\documentclass[twocolumn,aps,prl,notitlepage,superscriptaddress]{revtex4-1}
\usepackage{amsmath} 
\usepackage{amsthm} 
\usepackage{amssymb}	
\usepackage{graphicx} 
\usepackage[table]{xcolor}
\usepackage{comment}
\usepackage{dsfont}

\def\Tprep{T_p}
\def\Tramsey{T_s}
\def\Tread{T_r}

\newcommand{\ket}[1]{\left| #1 \right>} 
\newcommand{\bra}[1]{\left< #1 \right|} 

\begin{document}
\author{Soonwon~Choi}
\affiliation{Department of Physics, Harvard University, Cambridge, Massachusetts 02138, USA}
\author{Norman~Y.~Yao}
\affiliation{Department of Physics, University of California Berkeley, Berkeley, California 94720, USA}
\affiliation{Materials Science Division, Lawrence Berkeley National Laboratory, Berkeley, California 94720, USA}
\author{Mikhail~D.~Lukin}
\affiliation{Department of Physics, Harvard University, Cambridge, Massachusetts 02138, USA}

\date{\today}
\title{Quantum Metrology based on Strongly Correlated Matter}

\begin{abstract}
We propose and analyze a new method for quantum metrology based on stable non-equilibrium states of  quantum matter. Our approach utilizes quantum correlations stabilized by strong interactions and periodic driving.
As an example, we present an explicit protocol to perform Floquet enhanced measurements of an oscillating magnetic field in  Ising-interacting spin systems.
Our protocol  allows one to circumvent the interaction-induced decoherence associated with high density spin ensembles and is robust to the presence of noise and imperfections.
Applications to nanoscale magnetic sensing and precision measurements are discussed. 
\end{abstract}

\maketitle
The ability to interrogate a physical system and precisely measure its observables forms the basis of both  fundamental and applied sciences~\cite{Paola:2017RMP_sensing}.
While certain techniques are based on specially controlled individual particles~\cite{Diddams:2001da,Taylor:2008cp,Maze:2008cs}, in general, large ensembles can be used to enhance measurement sensitivity. 
For example, in the context of spectroscopy, collections of \emph{non-interacting} particles such as atoms, molecules, and electronic or nuclear spins are often used to maximize precision~\cite{Pham:2011dc,Bloom:2014OpticalLatticeClock,Budker:2014bt,Barry:2016gq,Glenn:2017kw}.
For an ensemble of $N$ uncorrelated two-level systems, the standard quantum limit (SQL) for measuring a small energy shift, scales as $\delta \omega \propto 1/\sqrt{N T_2 T}$, where $T_2$ is the relevant coherence time and $T$ is the total  measurement duration~\cite{Paola:2017RMP_sensing}. 
While this scaling suggests that increasing the number of particles always improves the signal to noise ratio, crucially, this argument does not capture the effect of inter-particle interactions. 
Above a certain density, these interactions fundamentally limit $T_2$ and thus the maximum achievable sensitivity. 
At its core, this limit arises from the fact that  interactions typically drive thermalization, wherein the system loses both its local coherences and any accumulated  spectroscopic signal.

Recent theoretical and experimental work has demonstrated that, under certain conditions, a many-body quantum system may evade rapid thermalization ~\cite{Basko:2006hh,nandkishore2015MBLreview,Schreiber:2015jt,Smith:2016cd,Kucsko:2016tn}.
Such intrinsically out-of-equilibrium systems can exhibit remarkably robust dynamical features that are forbidden in equilibrium~\cite{Lazarides:2015jd,Ponte:2015dc,Abanin:2016ev,Abanin:2015bc,Mori:2016wb,Khemani:2015gd,Else:2016gf,vonKeyserlingk:2016ev,Yao_dtc:2016wp}. 
One such example is the discrete time crystal (DTC)~\cite{Khemani:2015gd,Else:2016gf,vonKeyserlingk:2016ev,Yao_dtc:2016wp,Zhang:2016uw,Choi:2016wn,Ho:2017ea}, which is protected by an interplay between strong interactions and rapid periodic pulses~\cite{Else:2017kg}. The spatio-temporal ordering of the DTC phase is robust to arbitrary static perturbations and has been experimentally observed in both a trapped-ion quantum simulator~\cite{Zhang:2016uw} and a dipolar spin ensemble~\cite{Choi:2016wn}.

\begin{figure}
  \centering
  \includegraphics[width=0.45\textwidth]{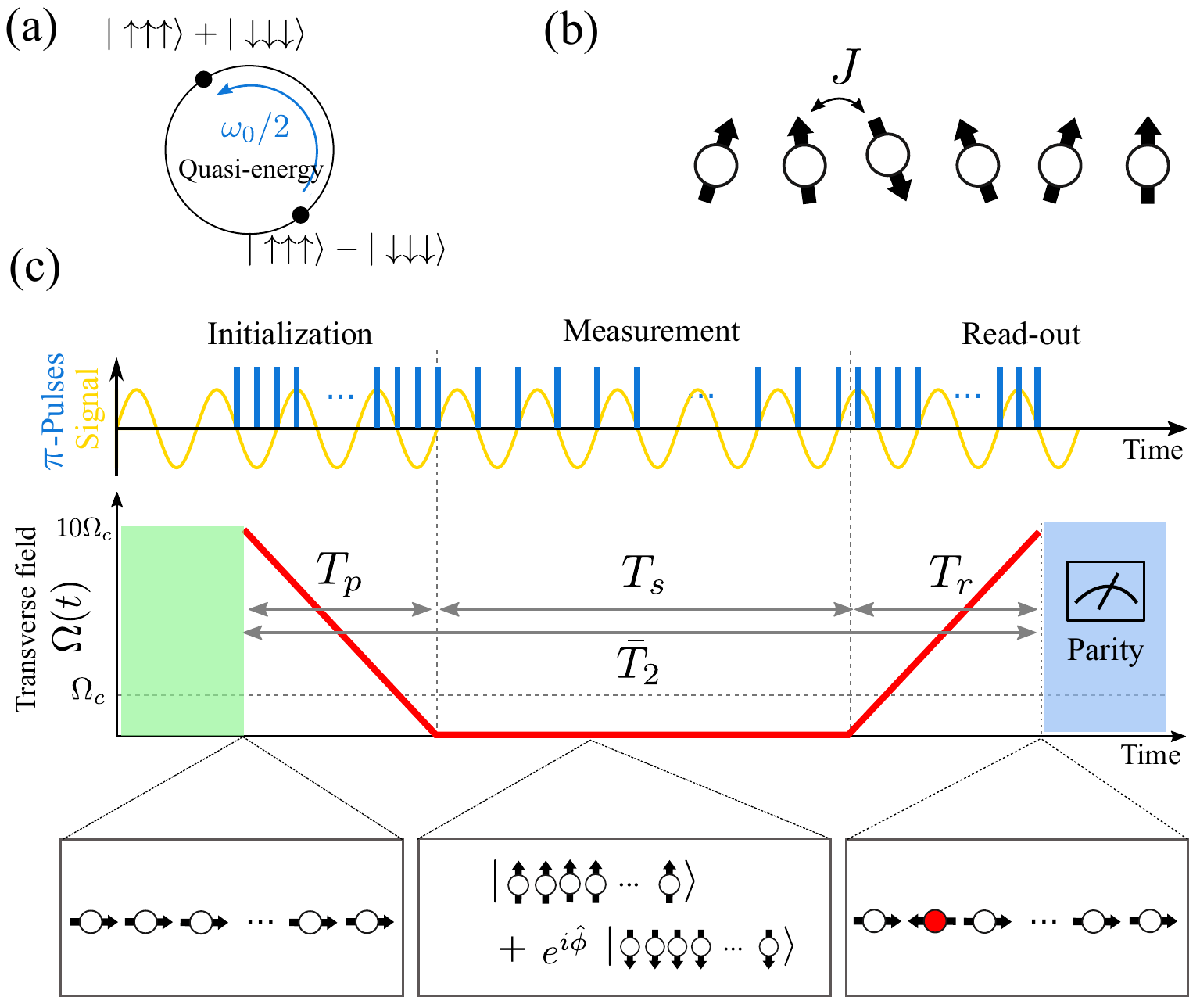}
  \caption{(a) Quasi-energy separation between two many-body entangled states. (b) Schematic for a spin chain with ferromagnetic interactions. (c) Sensing protocol. The protocol consists of three steps: initialization, measurement, and read-out. In each step, the transverse field $\Omega$ and the frequency of $\pi$-pulses are tuned. The signal strength is extracted from parity measurements.}
  \label{fig:schematic}
\end{figure}

In this Letter, we demonstrate that strongly interacting, non-equilibrium states of matter can be used to enhance quantum metrology.
In particular, we propose and analyze a class of protocols that allows one to circumvent limitations on the effective coherence time imposed by many-body interactions; rather, our protocols explicitly leverage interactions to develop additional quantum correlations leading to improved performances in both measurement sensitivity and bandwidth.
In the case of sensitivity, the enhancement partially arises from an ability to utilize a higher density of sensors, similar to prior studies \cite{Allred:2002bj,Deutsch:2010ky}, which utilize strong spin-exchange interactions to improve the spin lifetime \footnote{We note that our approach does not change the SQL sensitivity scaling. However, if the external noise which limits $T_2$ exhibits spatial correlations, it is well known that one can achieve an enhanced scaling with $N$ \cite{Paola:2017RMP_sensing}.}; on the other hand, our approach offers additional improvements in sensitivity and bandwidth, arising from an ability to prepare and utilize quantum correlated states.
The key idea is to engineer a Floquet system, where large quasi-energy gaps protect strongly-entangled states from static perturbations, while still ensuring their sensitivity to an oscillating signal [Fig~\ref{fig:schematic}(a)].
To this end, our approach can be understood as a generalization of spin-echo spectroscopy, where the states composing our effective two level system are in fact, entangled many-body states. 
More specifically, we employ periodic driving and strong interactions to stabilize Schr\"odinger's-cat-like states, which are typically extremely fragile against local perturbations in an equilibrium setting~\cite{Zanardi:2008ih,Macieszczak:2016gw,Skotiniotis:2015go,Frerot:2017uh}; thus, the essence of our approach is similar to the physics behind discrete time crystals~\cite{Khemani:2015gd,Else:2016gf,vonKeyserlingk:2016ev,Yao_dtc:2016wp,Else:2017kg,Ho:2017ea,Zhang:2016uw,Choi:2016wn}.
As specific examples, we analyze two techniques that allow for the precise measurement of AC magnetic fields in Floquet spin ensembles.
We demonstrate that these protocols are robust to imperfections in the external control parameters such as the strength and duration of the pulses. 
This robustness distinguishes our approach from prior studies that utilize either strongly correlated states or advanced dynamical decoupling techniques \cite{Zanardi:2008ih,Macieszczak:2016gw,Skotiniotis:2015go,Frerot:2017uh,Strobel:2014eg,Hosten:2016dj,Bohnet:2016ej,Aasi:2013jb,Hahn_echo:1950ge,deLange:2010ga,QDDproofLidar:2011aa,UDDproofJiang:2011aa,Choi:2017dynamical_engineering,NV_memory:2012Maurer,Lovchinsky:2016dq,Lovchinsky:2017gh}.

\emph{Sensing protocol.}---The central idea of our scheme can be understood by considering an ensemble of $N$ spin-1/2 particles in a $d$-dimensional array with ferromagnetic Ising interactions and a tunable transverse field $\Omega$ [Fig.~\ref{fig:schematic}(b)].
The spins are also coupled to a weak magnetic field signal, $B$, in the $\hat{z}$-direction  oscillating at frequency $\omega_s$.
The total Hamiltonian for the system is given by $\hat{H} = \hat{H}_0 + \hat{H}_\textrm{signal}$ with $\hat{H}_\textrm{signal} = B \sin(\omega_s t) \sum_i \hat{S}_i^z$ and 
\begin{align}
\hat{H}_0= - \sum_{ i,j} J_{ij} \hat{S}_i^z \hat{S}_{j}^z - \Omega \sum_{i}  \hat{S}_i^x ,
\end{align}
where $\hat{S}_i^\mu$ ($\mu \in \{x, y, z\}$) are spin-1/2 operators, and $J_{ij}>0$ is the interaction strength between spins at site-$i$ and $j$ with  characteristic strength $J\sim \sum_j J_{ij}$.
We envision that the spins are also subject to fast, global, periodic $\pi$-pulses, $\hat{P} \equiv \exp{[-i \pi \sum_i \hat{S}_i^x]}$, that rotate each spin around the $\hat{x}$ axis.
For $B=0$, the dynamics are driven by a Floquet unitary 
$\hat{U}_F = \hat{P} e^{-2\pi i \hat{H}_0 \tau}$,
where $\tau$ is the time-duration between $\pi$-pulses, setting the Floquet frequency, $\omega_0 = 2\pi / \tau$. Our goal is to measure the strength of the small magnetic field signal, $B$.
Our sensing protocol consists of three steps: (i) initialization, (ii) measurement, and (iii) read-out.
In each step, the transverse field $\Omega$ and the Floquet frequency $\omega_0$ are dynamically ramped to different values as indicated in Fig.~1(c).
In the initialization step, the spins are first polarized along the strong transverse field ($\Omega \gg J$), which is subsequently decreased to zero over time duration $T_p$.
During this process, $\omega_0$ is  sufficiently detuned  from $2\omega_s$ such that the effect of $\hat{H}_\textrm{signal}$ on the spin dynamics is negligible~\footnote{More specifically we require  $|\omega_0 - 2 \omega_s| \gg \Omega, J$.}.
In the measurement step, the Floquet frequency is adjusted to be resonant with the signal, $\omega_0 = 2\omega_s$, and the system evolves for a time duration $T_s$.
Finally, the initialization step is reversed over a time $T_r$ and each spin's  polarization is then measured along the $\hat{x}$ axis. 
These three steps must be completed within the relevant coherence time of the system, $\bar{T}_2 \ge T_p + T_s + T_r$ and will be repeated over a total integration time $T$.
As we will show, the magnetic field signal can be extracted from the average parity, $\langle \hat{P} \rangle \propto \langle \prod_i (2\hat{S}_i^x)\rangle$~\footnote{Note that the parity operator $P$ conicides with the unitary that globally rotates the spin ensemble by $\pi$ up to an unimportant complex phase.}.

In order to understand how sensitive the parity changes as a function of signal strength, we now analyze the Floquet dynamics in each of the three steps in detail.
During the initialization step, we utilize interactions to prepare a quantum state with strong spin-spin correlations. 
To understand the dynamics during  state preparation, we move into the so-called toggling frame, which rotates with every $\pi$-pulse, $\hat{P}$, by applying the unitary transformation $\hat{H} \mapsto \hat{P}^{-1} \hat{H} \hat{P}$~\cite{Slichter:2013wo}. 
In this frame, $\hat{H}_0$ remains invariant while $\hat{H}_\textrm{signal}$ changes sign during every Floquet period, modifying the time-dependence of the original signal to $B_\textrm{eff} (t) = B\sin(\omega_s t) \Theta (\omega_0 t /4\pi)$, where $\Theta(x)$ is a square function with unit period.
The dynamics of such a system are well approximated by an effective, quasi-static Hamiltonian~\cite{Kuwahara:2016dh,Abanin:2017hp}
\begin{align}
\hat{D} = \hat{H}_0+ \bar{B}_\textrm{eff} \sum_i \hat{S}_i^z 
\end{align}
where $\bar{B}_\textrm{eff}$ is the time-averaged signal strength and we have neglected a small correction of order $ \sim \mathcal{O}(B \Omega /\omega_s)$ \footnote{This description is valid up to an exponentially long time $\sim \exp{[\omega_0/\textrm{max}(\Omega,J)]}$, beyond which the system absorbs energy from the periodic driving and heats up to infinite temperature~\cite{Abanin:2015bc,Mori:2016wb,Kuwahara:2016dh,Abanin:2017hp}.}.
When $\omega_0$ is far-detuned from $2\omega_s$, $B_\textrm{eff} (t)$ rapidly oscillates with vanishing mean.
Our polarized initial state corresponds to the ground state of $\hat{D}$ for large $\Omega$.
As $\Omega$ is slowly decreased to zero, the system undergoes a phase transition from a paramagnet  to a ferromagnet with two degenerate ground states $\ket{G_\pm} = (\ket{\uparrow \uparrow \cdots \uparrow} \pm \ket{\downarrow \downarrow \cdots \downarrow})/\sqrt{2}$.
Crucially, during this process the effective Hamiltonian conserves parity and hence, for a sufficiently slow ramp, our initialization deterministically prepares the even parity Greenberger-Horne-Zeilinger (GHZ) state, $\ket{G_+}$, a well-known resource for quantum-enhanced metrology.

During the measurement step, tuning the Floquet frequency to $\omega_0 = 2\omega_s$ gives rise to a non-zero time-averaged signal strength $\bar{B}_\textrm{eff} =(2/\pi) B$, which resonantly couples the two degenerate ground states ($\ket{G_\pm}$), inducing coherent oscillations between them.
After time evolving for $T_s$, the system is in a superposition $\ket{G_\phi} = \cos(\phi) \ket{G_+} -i\sin(\phi) \ket{G_-}$, where $\phi = 2NBT_s$ is the collective phase accumulated by the spin ensemble during the measurement sequence.
This phase can be extracted from by measuring the parity $\langle \hat{P} \rangle = \cos(2\phi)$ in the paramagnetic phase, since $\ket{G_-}$ is adiabatically mapped to a single spin excitation, while $\ket{G_+}$ maps to zero excitations [see Fig.~1(c)].

Let us begin by analyzing the measurement sensitivity in the ideal case without external noise.
When our protocol is repeated  $k = T/\bar{T}_2$ times, the uncertainty in the phase is reduced to $\delta \phi \sim 1/\sqrt{k}$. Assuming a long measurement duration, $\bar{T}_2 \approx T_s \gg T_p,T_r$, the   sensitivity scales as:
\begin{align}
\delta B^{-1} \sim \delta \phi^{-1} NT_s \sim N \sqrt{\bar{T}_2 T} \label{eqn:heisenberg_scaling},
\end{align}
saturating the Heisenberg limit~\cite{Paola:2017RMP_sensing}. Note that the relevant coherence time here is determined by external noise at the probing frequency $\omega_s$ (and is not limited by interactions between the spins). 

A natural constraint for our protocol is the adiabatic preparation fidelity of the GHZ state.  The energy gap at the phase transition decreases with system size, which in turn requires a longer preparation step, and hence, a longer coherence time.
By crossing a phase transition in limited time, one necessarily creates a finite density of excitations.
However, even in this case, owing to ferromagnetic spin-spin correlations, our protocol can still achieve a sensitivity better than that of the SQL.
These correlations can be characterized by a  length scale $\xi$, where $\langle \hat{S}_i^z \hat{S}_j^z\rangle \sim e^{-|i-j|/\xi}$ and can be estimated from Kibble-Zurek scaling as $\xi \sim (JT_p)^{\nu/(1+z\nu)}$\,\cite{Zurek:2005cu}. Here,  $z$ and $\nu$ are the correlation-length and dynamical critical exponents of the transition \cite{Fisher:1972gh,Dutta:2001gu,Knap:2013jy,Fey:2016er,Maghrebi:2016bp,Elliott:1971hr,Pfeuty:1971ie,Friedman:1978jb}.

In this scenario, the initialization step prepares an even parity state of the form $\ket{\xi_+}=\left( \ket{\xi} + P\ket{\xi}\right)/\sqrt{2}$, where $\ket{\xi}$ is ferromagnetically ordered  with correlation length $\xi$.
[hat them]
The state accumulates a collective phase $\hat{\phi} = 4BT_s \sum_i \hat{S}_i^z $ during the measurement stage, leading to a  parity expectation value  $\langle \hat{P} \rangle = \bra{\xi} \cos (2\hat{\phi}) \ket{\xi}$.
For a weak signal, $\langle \hat{P} \rangle$ varies quadratically as: $1 - \langle \hat{P} \rangle \approx 2 \bra{\xi} \hat{\phi} \hat{\phi} \ket{\xi} \sim \xi^d N (BT_s)^2 $, which results in a sensitivity scaling 
\begin{align}
\label{eqn:scaling}
\delta B^{-1} \sim \sqrt{T / (\Tprep + \Tramsey + \Tread) } \sqrt{\xi^d N} \; \Tramsey.
\end{align}
Intuitively, this scaling be understood as follows.
The  state $\ket{\xi_+}$ can be viewed as multiple copies of a GHZ state with size $\sim\xi^d$.
While each GHZ state allows Heisenberg limited sensitivity $\sim \xi^d \sqrt{T\bar{T}_2}$, simultaneous measurements with all $N/\xi^d$ copies can further improve the signal to noise  by a factor $\sqrt{N/\xi^d}$, leading to the observed  scaling [Eq.~\eqref{eqn:scaling}].
When the correlation length approaches the system size, this scaling reaches the  limit in Eq.~\eqref{eqn:heisenberg_scaling}.

\begin{figure}
  \centering
  \includegraphics[width=0.41\textwidth]{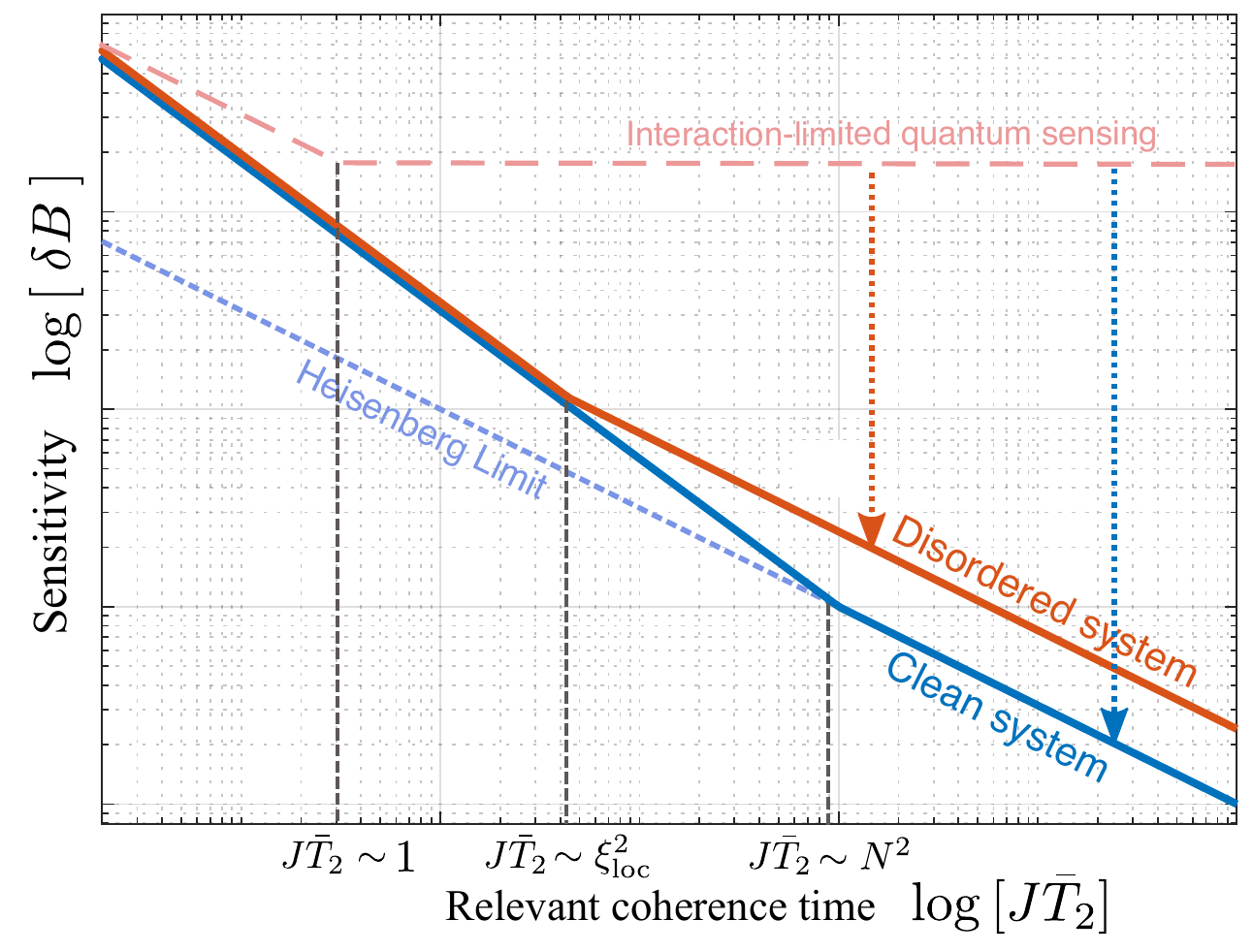}
  \caption{Sensitivity scaling $\delta B$ as a function of $\bar{T}_2$ for a given system size $N$ and total integration time $T$. As an example, we assumed a system of a spin chain with short-range interactions. Ideally, the sensitivity approaches Heisenberg limit when $\bar{T}_2$ is sufficiently long to prepare a GHZ state, e.g. $J\bar{T}_2 > N^2$. For a disordered system, this scaling is limited by the Anderson localization of quasi-particles with the localization length $\xi_\textrm{loc}$. In both cases, the protocol outperforms standard quantum limit, where the coherence time is limited by inter-particle interactions.}
  \label{fig:scaling_schematic}
\end{figure}
A few remarks are in order. 
First, given a limited coherence time, one should optimize the relative duration of each step.
This optimum is achieved when $\beta$\,$\equiv$\,$T_p/\bar{T}_2$\,$\simeq$\,$\,1- T_s/\bar{T}_2$\,=\,$\left(1 + 2 (\nu z+1)/d\nu\right)^{-1}$\,$\gg$\,$T_r/\bar{T}_2$.
For example, in one dimension with nearest neighbor interactions, the phase transition is characterized by exponents $\nu, z=1$,
 and the optimized sensitivity scales as 
\begin{align}
\delta B^{-1} \sim \sqrt{NT\bar{T}_2} (J \bar{T}_2)^{1/4}.
\end{align}
We note that this scaling improves upon the SQL by a factor  $\sim (J\bar{T}_2)^{1/4}$. 
Second, we emphasize that periodic $\pi$-pulses are essential to our protocol, since they suppress low frequency noise and prevent  changes to the parity of the spin ensemble. This protection originates from the quasi-energy gap between pairs of Floquet eigenstates with opposite parity [Fig.~1(a)].

\emph{Robustness.}---
We now turn to an analysis of our protocol in the presence of both imperfections and noise.
First, we consider quasi-static local perturbations $\epsilon \sum_i \delta \hat{H}_i$, which we decompose into  parity-preserving and parity-changing terms: $\delta \hat{H}_i = \delta \hat{H}_i^+ + \delta \hat{H}_i^-$ with $\delta \hat{H}_i^{\pm} \equiv (\delta \hat{H}_i \pm \hat{P} \delta \hat{H}_i \hat{P})/2$. 
The parity-preserving term, $\delta \hat{H}_i^+$, does not affect the nature of the phase transition nor  the sensitivity scaling of our protocol.
The parity-changing term, $\delta \hat{H}_i^-$, hinders both the state preparation and the  measurement fidelity of the magnetic field signal. 
However, this effect is parametrically suppressed by the presence of our periodic $\pi$-pulses, which effectively  ``echoes'' out this contribution to leading order. 
More specifically,  higher order corrections to the effective Hamiltonian appear only as $\sim \epsilon J /\omega_0$~\cite{Kuwahara:2016dh,Abanin:2017hp} and can be safely neglected assuming $\epsilon J / \omega_0 \ll  (\xi^d  \Tprep)^{-1}$ (initialization) and $\epsilon J /\omega_0 \ll B$ (measurement).

Second, we consider the presence of inhomogeneities in  $\delta \Omega_i$, $\delta J_i$, and $\delta \theta_i$.
Such inhomogeneities can lead to localization, which limits the maximum correlation length of the system.
In general, the localization length at the critical point scales as $\xi_\textrm{loc} \sim (W/J)^{-\mu}$, where $W$ is the disorder bandwidth of the coupling parameters and $\mu$ is the corresponding critical exponent. 
When the localization length $\xi_\textrm{loc}$ is shorter than the original correlation length $\xi \sim \sqrt{J \Tprep}$, one must reduce the state preparation time to $\Tprep^* = (W/J)^{-2\mu} / J $ (so  that more time can be allocated for the measurement step).
This leads to a  modified sensitivity scaling as summarized in Fig.~\ref{fig:scaling_schematic}
\footnote{Interestingly, the effect of disorder can be favorable during the measurement stage; while domain wall excitations can be mobile in the absence of disorder (in 1D), relatively weak disorder in $J$ may localize the excitations, allowing stable accumulation of phase information over long times. 
This effect is particularly relevant when the localization lengths during the measurement step is much shorter than that at the critical point, which is often satisfied in realistic systems, where the dominant source of disorder arises from random positioning of spins (disorder in $J$)~\cite{supp_info}.}.

Finally, we now consider the effect of external noise from the environment, which limits the coherence time, $\bar{T}_2$. Given a noise spectral density $S(\omega)\sim A_0^{1+\alpha}/\omega^{\alpha}$, the periodic $\pi$-pulses decouple the system from low frequency noise $\omega < \omega_s$, implying that the decoherence rate is determined by the noise density at the probe frequency, $S(\omega_s)$.
If the noise on each spin is independent, then the relevant coherence time of the entangled spin state is shortened to $\bar{T}_2\sim T_2^0 /\xi^d$, where $T_2^0$ is the lifetime of a single spin. 
In this case, the reduction of the coherence time off-sets any potential gain in the sensitivity in Eq.~\eqref{eqn:scaling}. This reduction is well-known and is in fact, fundamental for all methods that utilize entangled states for spectroscopy~\cite{Paola:2017RMP_sensing}. 
  We note, however, that our protocol still benefits from a shorter measurement duration $T_s$ (since the phase is accumulated $N$ times faster in $\ket{G_\phi}$), 
which provides a broader sensing bandwidth without compromising the sensitivity~\cite{supp_info}.
Finally, for solid-state spins, external noise often arises from nearby fluctuating dipole moments, which generates a spatially correlated  $S(\omega)$.
In this case, $\bar{T}_2$ can be significantly longer than $T_2^0 /\xi^d$ owing to  spatial averaging of the noise field in the collective phase $\hat{\phi} = 4BT_s \sum_i \hat{S}_i^z$, leading to an enhanced sensitivity ~\cite{supp_info}. 
\begin{figure}
  \centering
  \includegraphics[width=0.45\textwidth]{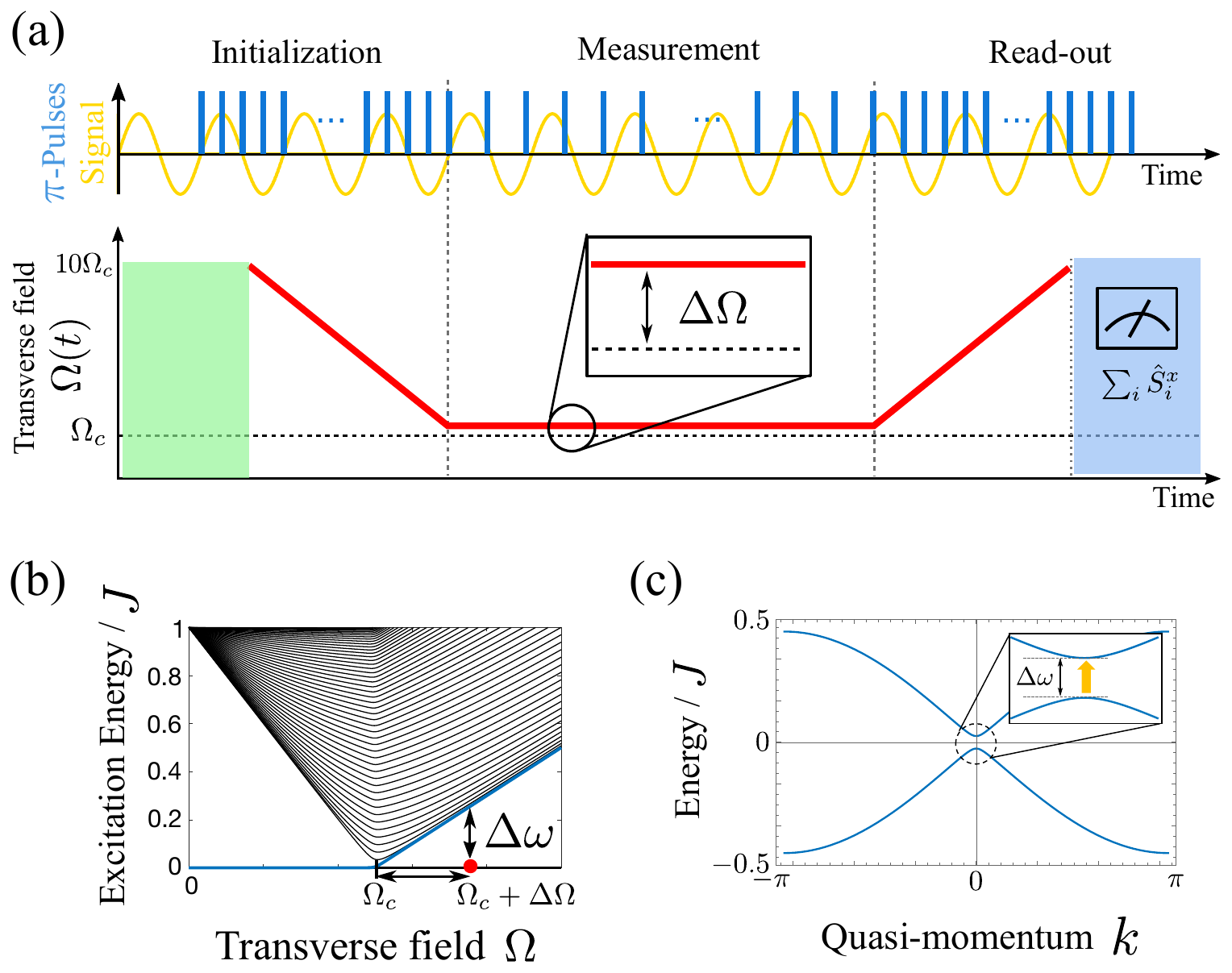}
  \caption{Sensing protocol without parity measurement. (a) Modified protocol. During the initialization, $\Omega$ does not cross the critical point. During the spectroscopy, $\pi$-pulse frequency is detuned from the signal frequency by $\Delta \omega$ (see main text). For read-out, the totally magnetization $\sum_i \hat{S}_i^x$ is measured. (b) Exemplary excitation spectrum of $\hat{H}_0$ (for a short-range interacting spin chain). In the vicinity of the critical point, the effective signal resonantly excites the system. (c) The dispersion relation of Bogoliubov quasi-particle excitations (for a short-range interacting spin chain). Low momentum modes are resonantly excited by the signal.}
  \label{fig:schematic2}
\end{figure}
\emph{Sensing protocol without parity measurements.}---
Parity measurements become challenging in an ensemble experiment where one lacks the ability to resolve individual spin projections.
To this end, we provide an alternative approach based upon measuring an extensively scaling observable.
Our modified protocol is shown in Fig.~\ref{fig:schematic2}(a).
During the initialization step, $\Omega$ is adiabatically decreased close to the critical point $\Omega= \Omega_c +\Delta \Omega $ without crossing the phase transition. 
Meanwhile, in the measurement step, rather than setting the Floquet frequency equal to $2\omega_s$, we now detune it by 
 $\Delta \omega \equiv \omega_s - \omega_0/2 $,  such that the magnetic field signal resonantly excites the system [Fig.~\ref{fig:schematic2}(b-c)]; in an experiment, this resonance condition would need to be calibrated.
Finally, $\Omega$ is slowly brought back to its original value, and the number of spin-flip excitations, $N_\textrm{e}$, now encodes the signal strength $B$ \footnote{For this protocol, excitations should not be created during the initialization or read-out steps. This condition can be estimated from the  Kibble-Zurek ``freezing point''  $\Delta \Omega \geq \Omega (JT_p)^{-1/(z\nu + 1)}$~\cite{Zurek:2005cu}.}.

The resonant magnetic field signal creates, on average, a single collective excitation within the correlation volume, $\xi^d$.  
The probability of creating such an excitation, $p\sim (\chi \xi^{d/2}BT_s)^{2}$, depends on the proximity to the critical point, which leads to the factor, $\chi \equiv (\Delta\Omega/\Omega)^{-\eta}$, where $\eta$ is the scaling dimension of the operator $\hat{S}_i^z$~\cite{sachdev2011quantum,Hauke:2016ht}.
Since there are $N/\xi^d$ correlated spin segments in the system, the average number of excitations $N_\textrm{e} \sim pN/\xi^d$, while its fluctuations $\delta {N_\textrm{e}}\,\sim\,\sqrt{p(1-p)N/\xi^d}$. This results in  a signal-to-noise ratio: $\partial_B  N_\textrm{e} / \delta N_\textrm{e}\sim \sqrt{N} T_s (JT_p)^{\eta/(z\nu + 1)}$. 
As before, when this procedure is repeated over a total duration $T$ with optimal $T_p$, the sensitivity scales as,
\begin{align}
\label{eqn:new_scheme_scaling}
\delta B^{-1} \sim \sqrt{NT\bar{T}_2} (J\bar{T}_2)^{\eta/(z\nu + 1)}.
\end{align}
For nearest neighbor interactions in 1D (Ising universality class), the scaling dimension is $\eta = 3/8$ and $\delta B^{-1} \sim \sqrt{NT\bar{T}_2} (J\bar{T}_2)^{3/16}$.~\cite{Bugrii:2001bh,Fonseca:2003ev,Essler:2009dj}

\emph{Implementations and Outlook.}---
Finally, we describe two potential platforms for realizing our protocol. 
First, we consider an AC magnetic field sensor using a 2D array of shallow implanted nitrogen-vacancy (NV) color centers in diamond \cite{Pham:2011dc,Barry:2016gq,Glenn:2017kw}.
The maximum sensitivity per unit area in this approach is limited by the dipolar interactions between the $S=1$ NV centers~\cite{Kucsko:2016tn}, which cannot be easily decoupled using conventional NMR techniques\,\cite{Waugh:1968im,Slichter:2013wo,Choi:2017dynamical_engineering}. 
Our protocol provides a way to circumvent this interaction-induced limitation and enable significant improvements to the sensitivity~\cite{supp_info}. 
A second platform for realizing our protocol is provided by nuclear spin ensembles in layered materials such as hexagonal boron-nitride or $^{13}$C enriched graphene. A particularly intriguing application of such systems includes the detection of time-varying signals resulting from weakly interacting massive particles such as axions~\cite{Budker:2014bt}.

Our scheme can also be extended along several directions. 
While we have focused on probing magnetic field signals,
similar methods can enable the detection of phase fluctuations in the external driving \cite{Diddams:2001da,Bloom:2014OpticalLatticeClock,Kolkowitz:2016gx}. 
Moreover, at present, our scheme enables the suppression of symmetry breaking perturbations at leading order via periodic $\pi$-pulses. 
An intriguing possibility is to extend such suppression to higher order corrections in the effective Hamiltonian. Indeed, in the limit of fast driving,  it has been shown that the system can exhibit an emergent symmetry up to exponentially long times ~\cite{Else:2017kg}.

\begin{acknowledgments}
The authors would like to thank W.\,W.~Ho, V.~Khemani, J.~Choi, H.~Zhou, E.~Altman, D.~Stamper-Kurn, and M.~Zaletel for useful discussions.
This work was supported through NSF, CUA, DOE, the Vannevar Bush Faculty Fellowship, the LDRD Program of LBNL, AFOSR MURI and Moore Foundation.
\end{acknowledgments}

\bibliography{refs}

\begin{thebibliography}{71}%
\makeatletter
\providecommand \@ifxundefined [1]{%
 \@ifx{#1\undefined}
}%
\providecommand \@ifnum [1]{%
 \ifnum #1\expandafter \@firstoftwo
 \else \expandafter \@secondoftwo
 \fi
}%
\providecommand \@ifx [1]{%
 \ifx #1\expandafter \@firstoftwo
 \else \expandafter \@secondoftwo
 \fi
}%
\providecommand \natexlab [1]{#1}%
\providecommand \enquote  [1]{``#1''}%
\providecommand \bibnamefont  [1]{#1}%
\providecommand \bibfnamefont [1]{#1}%
\providecommand \citenamefont [1]{#1}%
\providecommand \href@noop [0]{\@secondoftwo}%
\providecommand \href [0]{\begingroup \@sanitize@url \@href}%
\providecommand \@href[1]{\@@startlink{#1}\@@href}%
\providecommand \@@href[1]{\endgroup#1\@@endlink}%
\providecommand \@sanitize@url [0]{\catcode `\\12\catcode `\$12\catcode
  `\&12\catcode `\#12\catcode `\^12\catcode `\_12\catcode `\%12\relax}%
\providecommand \@@startlink[1]{}%
\providecommand \@@endlink[0]{}%
\providecommand \url  [0]{\begingroup\@sanitize@url \@url }%
\providecommand \@url [1]{\endgroup\@href {#1}{\urlprefix }}%
\providecommand \urlprefix  [0]{URL }%
\providecommand \Eprint [0]{\href }%
\providecommand \doibase [0]{http://dx.doi.org/}%
\providecommand \selectlanguage [0]{\@gobble}%
\providecommand \bibinfo  [0]{\@secondoftwo}%
\providecommand \bibfield  [0]{\@secondoftwo}%
\providecommand \translation [1]{[#1]}%
\providecommand \BibitemOpen [0]{}%
\providecommand \bibitemStop [0]{}%
\providecommand \bibitemNoStop [0]{.\EOS\space}%
\providecommand \EOS [0]{\spacefactor3000\relax}%
\providecommand \BibitemShut  [1]{\csname bibitem#1\endcsname}%
\let\auto@bib@innerbib\@empty
\bibitem [{\citenamefont {Degen}\ \emph {et~al.}(2017)\citenamefont {Degen},
  \citenamefont {Reinhard},\ and\ \citenamefont
  {Cappellaro}}]{Paola:2017RMP_sensing}%
  \BibitemOpen
  \bibfield  {author} {\bibinfo {author} {\bibfnamefont {C.~L.}\ \bibnamefont
  {Degen}}, \bibinfo {author} {\bibfnamefont {F.}~\bibnamefont {Reinhard}}, \
  and\ \bibinfo {author} {\bibfnamefont {P.}~\bibnamefont {Cappellaro}},\
  }\href@noop {} {\bibfield  {journal} {\bibinfo  {journal} {Rev. Mod. Phys.}\
  }\textbf {\bibinfo {volume} {89}},\ \bibinfo {pages} {035002} (\bibinfo
  {year} {2017})}\BibitemShut {NoStop}%
\bibitem [{\citenamefont {Diddams}\ \emph {et~al.}(2001)\citenamefont
  {Diddams}, \citenamefont {Udem}, \citenamefont {Bergquist}, \citenamefont
  {Curtis}, \citenamefont {Drullinger}, \citenamefont {Hollberg}, \citenamefont
  {Itano}, \citenamefont {Lee}, \citenamefont {Oates}, \citenamefont {Vogel},\
  and\ \citenamefont {Wineland}}]{Diddams:2001da}%
  \BibitemOpen
  \bibfield  {author} {\bibinfo {author} {\bibfnamefont {S.~A.}\ \bibnamefont
  {Diddams}}, \bibinfo {author} {\bibfnamefont {T.}~\bibnamefont {Udem}},
  \bibinfo {author} {\bibfnamefont {J.~C.}\ \bibnamefont {Bergquist}}, \bibinfo
  {author} {\bibfnamefont {E.~A.}\ \bibnamefont {Curtis}}, \bibinfo {author}
  {\bibfnamefont {R.~E.}\ \bibnamefont {Drullinger}}, \bibinfo {author}
  {\bibfnamefont {L.}~\bibnamefont {Hollberg}}, \bibinfo {author}
  {\bibfnamefont {W.~M.}\ \bibnamefont {Itano}}, \bibinfo {author}
  {\bibfnamefont {W.~D.}\ \bibnamefont {Lee}}, \bibinfo {author} {\bibfnamefont
  {C.~W.}\ \bibnamefont {Oates}}, \bibinfo {author} {\bibfnamefont {K.~R.}\
  \bibnamefont {Vogel}}, \ and\ \bibinfo {author} {\bibfnamefont {D.~J.}\
  \bibnamefont {Wineland}},\ }\href@noop {} {\bibfield  {journal} {\bibinfo
  {journal} {Science}\ }\textbf {\bibinfo {volume} {293}},\ \bibinfo {pages}
  {825} (\bibinfo {year} {2001})}\BibitemShut {NoStop}%
\bibitem [{\citenamefont {Taylor}\ \emph {et~al.}(2008)\citenamefont {Taylor},
  \citenamefont {Cappellaro}, \citenamefont {Childress}, \citenamefont {Jiang},
  \citenamefont {Budker}, \citenamefont {Hemmer}, \citenamefont {Yacoby},
  \citenamefont {Walsworth},\ and\ \citenamefont {Lukin}}]{Taylor:2008cp}%
  \BibitemOpen
  \bibfield  {author} {\bibinfo {author} {\bibfnamefont {J.~M.}\ \bibnamefont
  {Taylor}}, \bibinfo {author} {\bibfnamefont {P.}~\bibnamefont {Cappellaro}},
  \bibinfo {author} {\bibfnamefont {L.}~\bibnamefont {Childress}}, \bibinfo
  {author} {\bibfnamefont {L.}~\bibnamefont {Jiang}}, \bibinfo {author}
  {\bibfnamefont {D.}~\bibnamefont {Budker}}, \bibinfo {author} {\bibfnamefont
  {P.~R.}\ \bibnamefont {Hemmer}}, \bibinfo {author} {\bibfnamefont
  {A.}~\bibnamefont {Yacoby}}, \bibinfo {author} {\bibfnamefont
  {R.}~\bibnamefont {Walsworth}}, \ and\ \bibinfo {author} {\bibfnamefont
  {M.~D.}\ \bibnamefont {Lukin}},\ }\href@noop {} {\bibfield  {journal}
  {\bibinfo  {journal} {Nature Physics}\ }\textbf {\bibinfo {volume} {4}},\
  \bibinfo {pages} {810} (\bibinfo {year} {2008})}\BibitemShut {NoStop}%
\bibitem [{\citenamefont {Maze}\ \emph {et~al.}(2008)\citenamefont {Maze},
  \citenamefont {Stanwix}, \citenamefont {Hodges}, \citenamefont {Hong},
  \citenamefont {Taylor}, \citenamefont {Cappellaro}, \citenamefont {Jiang},
  \citenamefont {Dutt}, \citenamefont {Togan}, \citenamefont {Zibrov},
  \citenamefont {Yacoby}, \citenamefont {Walsworth},\ and\ \citenamefont
  {Lukin}}]{Maze:2008cs}%
  \BibitemOpen
  \bibfield  {author} {\bibinfo {author} {\bibfnamefont {J.~R.}\ \bibnamefont
  {Maze}}, \bibinfo {author} {\bibfnamefont {P.~L.}\ \bibnamefont {Stanwix}},
  \bibinfo {author} {\bibfnamefont {J.~S.}\ \bibnamefont {Hodges}}, \bibinfo
  {author} {\bibfnamefont {S.}~\bibnamefont {Hong}}, \bibinfo {author}
  {\bibfnamefont {J.~M.}\ \bibnamefont {Taylor}}, \bibinfo {author}
  {\bibfnamefont {P.}~\bibnamefont {Cappellaro}}, \bibinfo {author}
  {\bibfnamefont {L.}~\bibnamefont {Jiang}}, \bibinfo {author} {\bibfnamefont
  {M.~V.~G.}\ \bibnamefont {Dutt}}, \bibinfo {author} {\bibfnamefont
  {E.}~\bibnamefont {Togan}}, \bibinfo {author} {\bibfnamefont {A.~S.}\
  \bibnamefont {Zibrov}}, \bibinfo {author} {\bibfnamefont {A.}~\bibnamefont
  {Yacoby}}, \bibinfo {author} {\bibfnamefont {R.~L.}\ \bibnamefont
  {Walsworth}}, \ and\ \bibinfo {author} {\bibfnamefont {M.~D.}\ \bibnamefont
  {Lukin}},\ }\href@noop {} {\bibfield  {journal} {\bibinfo  {journal}
  {Nature}\ }\textbf {\bibinfo {volume} {455}},\ \bibinfo {pages} {644}
  (\bibinfo {year} {2008})}\BibitemShut {NoStop}%
\bibitem [{\citenamefont {Pham}\ \emph {et~al.}(2011)\citenamefont {Pham},
  \citenamefont {Le~Sage}, \citenamefont {Stanwix}, \citenamefont {Yeung},
  \citenamefont {Glenn}, \citenamefont {Trifonov}, \citenamefont {Cappellaro},
  \citenamefont {Hemmer}, \citenamefont {Lukin}, \citenamefont {Park},
  \citenamefont {Yacoby},\ and\ \citenamefont {Walsworth}}]{Pham:2011dc}%
  \BibitemOpen
  \bibfield  {author} {\bibinfo {author} {\bibfnamefont {L.~M.}\ \bibnamefont
  {Pham}}, \bibinfo {author} {\bibfnamefont {D.}~\bibnamefont {Le~Sage}},
  \bibinfo {author} {\bibfnamefont {P.~L.}\ \bibnamefont {Stanwix}}, \bibinfo
  {author} {\bibfnamefont {T.~K.}\ \bibnamefont {Yeung}}, \bibinfo {author}
  {\bibfnamefont {D.}~\bibnamefont {Glenn}}, \bibinfo {author} {\bibfnamefont
  {A.}~\bibnamefont {Trifonov}}, \bibinfo {author} {\bibfnamefont
  {P.}~\bibnamefont {Cappellaro}}, \bibinfo {author} {\bibfnamefont {P.~R.}\
  \bibnamefont {Hemmer}}, \bibinfo {author} {\bibfnamefont {M.~D.}\
  \bibnamefont {Lukin}}, \bibinfo {author} {\bibfnamefont {H.}~\bibnamefont
  {Park}}, \bibinfo {author} {\bibfnamefont {A.}~\bibnamefont {Yacoby}}, \ and\
  \bibinfo {author} {\bibfnamefont {R.~L.}\ \bibnamefont {Walsworth}},\
  }\href@noop {} {\bibfield  {journal} {\bibinfo  {journal} {New Journal of
  Physics}\ }\textbf {\bibinfo {volume} {13}},\ \bibinfo {pages} {045021}
  (\bibinfo {year} {2011})}\BibitemShut {NoStop}%
\bibitem [{Blo(2014)}]{Bloom:2014OpticalLatticeClock}%
  \BibitemOpen
  \href@noop {} {\ \textbf {\bibinfo {volume} {506}},\ \bibinfo {pages} {71}
  (\bibinfo {year} {2014})}\BibitemShut {NoStop}%
\bibitem [{\citenamefont {Budker}\ \emph {et~al.}(2014)\citenamefont {Budker},
  \citenamefont {Graham}, \citenamefont {Ledbetter}, \citenamefont
  {Rajendran},\ and\ \citenamefont {Sushkov}}]{Budker:2014bt}%
  \BibitemOpen
  \bibfield  {author} {\bibinfo {author} {\bibfnamefont {D.}~\bibnamefont
  {Budker}}, \bibinfo {author} {\bibfnamefont {P.~W.}\ \bibnamefont {Graham}},
  \bibinfo {author} {\bibfnamefont {M.}~\bibnamefont {Ledbetter}}, \bibinfo
  {author} {\bibfnamefont {S.}~\bibnamefont {Rajendran}}, \ and\ \bibinfo
  {author} {\bibfnamefont {A.~O.}\ \bibnamefont {Sushkov}},\ }\href@noop {}
  {\bibfield  {journal} {\bibinfo  {journal} {Physical Review X}\ }\textbf
  {\bibinfo {volume} {4}},\ \bibinfo {pages} {021030} (\bibinfo {year}
  {2014})}\BibitemShut {NoStop}%
\bibitem [{\citenamefont {Barry}\ \emph {et~al.}(2016)\citenamefont {Barry},
  \citenamefont {Turner}, \citenamefont {Schloss}, \citenamefont {Glenn},
  \citenamefont {Song}, \citenamefont {Lukin}, \citenamefont {Park},\ and\
  \citenamefont {Walsworth}}]{Barry:2016gq}%
  \BibitemOpen
  \bibfield  {author} {\bibinfo {author} {\bibfnamefont {J.~F.}\ \bibnamefont
  {Barry}}, \bibinfo {author} {\bibfnamefont {M.~J.}\ \bibnamefont {Turner}},
  \bibinfo {author} {\bibfnamefont {J.~M.}\ \bibnamefont {Schloss}}, \bibinfo
  {author} {\bibfnamefont {D.~R.}\ \bibnamefont {Glenn}}, \bibinfo {author}
  {\bibfnamefont {Y.}~\bibnamefont {Song}}, \bibinfo {author} {\bibfnamefont
  {M.~D.}\ \bibnamefont {Lukin}}, \bibinfo {author} {\bibfnamefont
  {H.}~\bibnamefont {Park}}, \ and\ \bibinfo {author} {\bibfnamefont {R.~L.}\
  \bibnamefont {Walsworth}},\ }\href@noop {} {\bibfield  {journal} {\bibinfo
  {journal} {Proceedings of the National Academy of Sciences of the United
  States of America}\ }\textbf {\bibinfo {volume} {113}},\ \bibinfo {pages}
  {14133} (\bibinfo {year} {2016})}\BibitemShut {NoStop}%
\bibitem [{\citenamefont {Glenn}\ \emph {et~al.}(2017)\citenamefont {Glenn},
  \citenamefont {Fu}, \citenamefont {Kehayias}, \citenamefont {Le~Sage},
  \citenamefont {Lima}, \citenamefont {Weiss},\ and\ \citenamefont
  {Walsworth}}]{Glenn:2017kw}%
  \BibitemOpen
  \bibfield  {author} {\bibinfo {author} {\bibfnamefont {D.~R.}\ \bibnamefont
  {Glenn}}, \bibinfo {author} {\bibfnamefont {R.~R.}\ \bibnamefont {Fu}},
  \bibinfo {author} {\bibfnamefont {P.}~\bibnamefont {Kehayias}}, \bibinfo
  {author} {\bibfnamefont {D.}~\bibnamefont {Le~Sage}}, \bibinfo {author}
  {\bibfnamefont {E.~A.}\ \bibnamefont {Lima}}, \bibinfo {author}
  {\bibfnamefont {B.~P.}\ \bibnamefont {Weiss}}, \ and\ \bibinfo {author}
  {\bibfnamefont {R.~L.}\ \bibnamefont {Walsworth}},\ }\href@noop {} {\bibfield
   {journal} {\bibinfo  {journal} {Geochemistry, Geophysics, Geosystems}\
  }\textbf {\bibinfo {volume} {18}},\ \bibinfo {pages} {3254} (\bibinfo {year}
  {2017})}\BibitemShut {NoStop}%
\bibitem [{\citenamefont {Basko}\ \emph {et~al.}(2006)\citenamefont {Basko},
  \citenamefont {Aleiner},\ and\ \citenamefont {Altshuler}}]{Basko:2006hh}%
  \BibitemOpen
  \bibfield  {author} {\bibinfo {author} {\bibfnamefont {D.~M.}\ \bibnamefont
  {Basko}}, \bibinfo {author} {\bibfnamefont {I.~L.}\ \bibnamefont {Aleiner}},
  \ and\ \bibinfo {author} {\bibfnamefont {B.~L.}\ \bibnamefont {Altshuler}},\
  }\href@noop {} {\bibfield  {journal} {\bibinfo  {journal} {Annals of
  Physics}\ }\textbf {\bibinfo {volume} {321}},\ \bibinfo {pages} {1126}
  (\bibinfo {year} {2006})}\BibitemShut {NoStop}%
\bibitem [{\citenamefont {Nandkishore}\ and\ \citenamefont
  {Huse}(2015)}]{nandkishore2015MBLreview}%
  \BibitemOpen
  \bibfield  {author} {\bibinfo {author} {\bibfnamefont {R.}~\bibnamefont
  {Nandkishore}}\ and\ \bibinfo {author} {\bibfnamefont {D.~A.}\ \bibnamefont
  {Huse}},\ }\href@noop {} {\bibfield  {journal} {\bibinfo  {journal} {Annual
  Review of Condensed Matter Physics}\ }\textbf {\bibinfo {volume} {6}},\
  \bibinfo {pages} {15} (\bibinfo {year} {2015})}\BibitemShut {NoStop}%
\bibitem [{\citenamefont {Schreiber}\ \emph {et~al.}(2015)\citenamefont
  {Schreiber}, \citenamefont {Hodgman}, \citenamefont {Bordia}, \citenamefont
  {L{\"u}schen}, \citenamefont {Fischer}, \citenamefont {Vosk}, \citenamefont
  {Altman}, \citenamefont {Schneider},\ and\ \citenamefont
  {Bloch}}]{Schreiber:2015jt}%
  \BibitemOpen
  \bibfield  {author} {\bibinfo {author} {\bibfnamefont {M.}~\bibnamefont
  {Schreiber}}, \bibinfo {author} {\bibfnamefont {S.~S.}\ \bibnamefont
  {Hodgman}}, \bibinfo {author} {\bibfnamefont {P.}~\bibnamefont {Bordia}},
  \bibinfo {author} {\bibfnamefont {H.~P.}\ \bibnamefont {L{\"u}schen}},
  \bibinfo {author} {\bibfnamefont {M.~H.}\ \bibnamefont {Fischer}}, \bibinfo
  {author} {\bibfnamefont {R.}~\bibnamefont {Vosk}}, \bibinfo {author}
  {\bibfnamefont {E.}~\bibnamefont {Altman}}, \bibinfo {author} {\bibfnamefont
  {U.}~\bibnamefont {Schneider}}, \ and\ \bibinfo {author} {\bibfnamefont
  {I.}~\bibnamefont {Bloch}},\ }\href@noop {} {\bibfield  {journal} {\bibinfo
  {journal} {Science}\ }\textbf {\bibinfo {volume} {349}},\ \bibinfo {pages}
  {842} (\bibinfo {year} {2015})}\BibitemShut {NoStop}%
\bibitem [{\citenamefont {Smith}\ \emph {et~al.}(2016)\citenamefont {Smith},
  \citenamefont {Lee}, \citenamefont {Richerme}, \citenamefont {Neyenhuis},
  \citenamefont {Hess}, \citenamefont {Hauke}, \citenamefont {Heyl},
  \citenamefont {Huse},\ and\ \citenamefont {Monroe}}]{Smith:2016cd}%
  \BibitemOpen
  \bibfield  {author} {\bibinfo {author} {\bibfnamefont {J.}~\bibnamefont
  {Smith}}, \bibinfo {author} {\bibfnamefont {A.}~\bibnamefont {Lee}}, \bibinfo
  {author} {\bibfnamefont {P.}~\bibnamefont {Richerme}}, \bibinfo {author}
  {\bibfnamefont {B.}~\bibnamefont {Neyenhuis}}, \bibinfo {author}
  {\bibfnamefont {P.~W.}\ \bibnamefont {Hess}}, \bibinfo {author}
  {\bibfnamefont {P.}~\bibnamefont {Hauke}}, \bibinfo {author} {\bibfnamefont
  {M.}~\bibnamefont {Heyl}}, \bibinfo {author} {\bibfnamefont {D.~A.}\
  \bibnamefont {Huse}}, \ and\ \bibinfo {author} {\bibfnamefont
  {C.}~\bibnamefont {Monroe}},\ }\href@noop {} {\bibfield  {journal} {\bibinfo
  {journal} {Nature Physics}\ }\textbf {\bibinfo {volume} {12}},\ \bibinfo
  {pages} {907} (\bibinfo {year} {2016})}\BibitemShut {NoStop}%
\bibitem [{\citenamefont {Kucsko}\ \emph {et~al.}(2016)\citenamefont {Kucsko},
  \citenamefont {Choi}, \citenamefont {Choi}, \citenamefont {Maurer},
  \citenamefont {Sumiya}, \citenamefont {Onoda}, \citenamefont {Isoya},
  \citenamefont {Jelezko}, \citenamefont {Demler}, \citenamefont {Yao},\ and\
  \citenamefont {Lukin}}]{Kucsko:2016tn}%
  \BibitemOpen
  \bibfield  {author} {\bibinfo {author} {\bibfnamefont {G.}~\bibnamefont
  {Kucsko}}, \bibinfo {author} {\bibfnamefont {S.}~\bibnamefont {Choi}},
  \bibinfo {author} {\bibfnamefont {J.}~\bibnamefont {Choi}}, \bibinfo {author}
  {\bibfnamefont {P.~C.}\ \bibnamefont {Maurer}}, \bibinfo {author}
  {\bibfnamefont {H.}~\bibnamefont {Sumiya}}, \bibinfo {author} {\bibfnamefont
  {S.}~\bibnamefont {Onoda}}, \bibinfo {author} {\bibfnamefont
  {J.}~\bibnamefont {Isoya}}, \bibinfo {author} {\bibfnamefont
  {F.}~\bibnamefont {Jelezko}}, \bibinfo {author} {\bibfnamefont
  {E.}~\bibnamefont {Demler}}, \bibinfo {author} {\bibfnamefont {N.~Y.}\
  \bibnamefont {Yao}}, \ and\ \bibinfo {author} {\bibfnamefont {M.~D.}\
  \bibnamefont {Lukin}},\ }\href@noop {} {\  (\bibinfo {year}
  {2016})}\BibitemShut {NoStop}%
\bibitem [{\citenamefont {Lazarides}\ \emph {et~al.}(2015)\citenamefont
  {Lazarides}, \citenamefont {Das},\ and\ \citenamefont
  {Moessner}}]{Lazarides:2015jd}%
  \BibitemOpen
  \bibfield  {author} {\bibinfo {author} {\bibfnamefont {A.}~\bibnamefont
  {Lazarides}}, \bibinfo {author} {\bibfnamefont {A.}~\bibnamefont {Das}}, \
  and\ \bibinfo {author} {\bibfnamefont {R.}~\bibnamefont {Moessner}},\
  }\href@noop {} {\bibfield  {journal} {\bibinfo  {journal} {Physical Review
  Letters}\ }\textbf {\bibinfo {volume} {115}},\ \bibinfo {pages} {030402}
  (\bibinfo {year} {2015})}\BibitemShut {NoStop}%
\bibitem [{\citenamefont {Ponte}\ \emph {et~al.}(2015)\citenamefont {Ponte},
  \citenamefont {Papi{\'c}}, \citenamefont {Huveneers},\ and\ \citenamefont
  {Abanin}}]{Ponte:2015dc}%
  \BibitemOpen
  \bibfield  {author} {\bibinfo {author} {\bibfnamefont {P.}~\bibnamefont
  {Ponte}}, \bibinfo {author} {\bibfnamefont {Z.}~\bibnamefont {Papi{\'c}}},
  \bibinfo {author} {\bibfnamefont {F.}~\bibnamefont {Huveneers}}, \ and\
  \bibinfo {author} {\bibfnamefont {D.~A.}\ \bibnamefont {Abanin}},\
  }\href@noop {} {\bibfield  {journal} {\bibinfo  {journal} {Physical Review
  Letters}\ }\textbf {\bibinfo {volume} {114}},\ \bibinfo {pages} {140401}
  (\bibinfo {year} {2015})}\BibitemShut {NoStop}%
\bibitem [{\citenamefont {Abanin}\ \emph {et~al.}(2016)\citenamefont {Abanin},
  \citenamefont {De~Roeck},\ and\ \citenamefont {Huveneers}}]{Abanin:2016ev}%
  \BibitemOpen
  \bibfield  {author} {\bibinfo {author} {\bibfnamefont {D.~A.}\ \bibnamefont
  {Abanin}}, \bibinfo {author} {\bibfnamefont {W.}~\bibnamefont {De~Roeck}}, \
  and\ \bibinfo {author} {\bibfnamefont {F.}~\bibnamefont {Huveneers}},\
  }\href@noop {} {\bibfield  {journal} {\bibinfo  {journal} {Annals of
  Physics}\ }\textbf {\bibinfo {volume} {372}},\ \bibinfo {pages} {1} (\bibinfo
  {year} {2016})}\BibitemShut {NoStop}%
\bibitem [{\citenamefont {Abanin}\ \emph {et~al.}(2015)\citenamefont {Abanin},
  \citenamefont {De~Roeck},\ and\ \citenamefont {Huveneers}}]{Abanin:2015bc}%
  \BibitemOpen
  \bibfield  {author} {\bibinfo {author} {\bibfnamefont {D.~A.}\ \bibnamefont
  {Abanin}}, \bibinfo {author} {\bibfnamefont {W.}~\bibnamefont {De~Roeck}}, \
  and\ \bibinfo {author} {\bibfnamefont {F.}~\bibnamefont {Huveneers}},\
  }\href@noop {} {\bibfield  {journal} {\bibinfo  {journal} {Physical Review
  Letters}\ }\textbf {\bibinfo {volume} {115}},\ \bibinfo {pages} {256803}
  (\bibinfo {year} {2015})}\BibitemShut {NoStop}%
\bibitem [{\citenamefont {Mori}\ \emph {et~al.}(2016)\citenamefont {Mori},
  \citenamefont {Kuwahara},\ and\ \citenamefont {Saito}}]{Mori:2016wb}%
  \BibitemOpen
  \bibfield  {author} {\bibinfo {author} {\bibfnamefont {T.}~\bibnamefont
  {Mori}}, \bibinfo {author} {\bibfnamefont {T.}~\bibnamefont {Kuwahara}}, \
  and\ \bibinfo {author} {\bibfnamefont {K.}~\bibnamefont {Saito}},\
  }\href@noop {} {\bibfield  {journal} {\bibinfo  {journal} {Physical Review
  Letters}\ }\textbf {\bibinfo {volume} {116}},\ \bibinfo {pages} {120401}
  (\bibinfo {year} {2016})}\BibitemShut {NoStop}%
\bibitem [{\citenamefont {Khemani}\ \emph {et~al.}(2016)\citenamefont
  {Khemani}, \citenamefont {Lazarides}, \citenamefont {Moessner},\ and\
  \citenamefont {Sondhi}}]{Khemani:2015gd}%
  \BibitemOpen
  \bibfield  {author} {\bibinfo {author} {\bibfnamefont {V.}~\bibnamefont
  {Khemani}}, \bibinfo {author} {\bibfnamefont {A.}~\bibnamefont {Lazarides}},
  \bibinfo {author} {\bibfnamefont {R.}~\bibnamefont {Moessner}}, \ and\
  \bibinfo {author} {\bibfnamefont {S.~L.}\ \bibnamefont {Sondhi}},\
  }\href@noop {} {\bibfield  {journal} {\bibinfo  {journal} {Physical Review
  Letters}\ }\textbf {\bibinfo {volume} {116}},\ \bibinfo {pages} {250401}
  (\bibinfo {year} {2016})}\BibitemShut {NoStop}%
\bibitem [{\citenamefont {Else}\ \emph {et~al.}(2016)\citenamefont {Else},
  \citenamefont {Bauer},\ and\ \citenamefont {Nayak}}]{Else:2016gf}%
  \BibitemOpen
  \bibfield  {author} {\bibinfo {author} {\bibfnamefont {D.~V.}\ \bibnamefont
  {Else}}, \bibinfo {author} {\bibfnamefont {B.}~\bibnamefont {Bauer}}, \ and\
  \bibinfo {author} {\bibfnamefont {C.}~\bibnamefont {Nayak}},\ }\href@noop {}
  {\bibfield  {journal} {\bibinfo  {journal} {Physical Review Letters}\
  }\textbf {\bibinfo {volume} {117}},\ \bibinfo {pages} {090402} (\bibinfo
  {year} {2016})}\BibitemShut {NoStop}%
\bibitem [{\citenamefont {von Keyserlingk}\ \emph {et~al.}(2016)\citenamefont
  {von Keyserlingk}, \citenamefont {Khemani},\ and\ \citenamefont
  {Sondhi}}]{vonKeyserlingk:2016ev}%
  \BibitemOpen
  \bibfield  {author} {\bibinfo {author} {\bibfnamefont {C.~W.}\ \bibnamefont
  {von Keyserlingk}}, \bibinfo {author} {\bibfnamefont {V.}~\bibnamefont
  {Khemani}}, \ and\ \bibinfo {author} {\bibfnamefont {S.~L.}\ \bibnamefont
  {Sondhi}},\ }\href@noop {} {\bibfield  {journal} {\bibinfo  {journal}
  {Physical Review B}\ }\textbf {\bibinfo {volume} {94}},\ \bibinfo {pages}
  {085112} (\bibinfo {year} {2016})}\BibitemShut {NoStop}%
\bibitem [{\citenamefont {Yao}\ \emph {et~al.}(2017)\citenamefont {Yao},
  \citenamefont {Potter}, \citenamefont {Potirniche},\ and\ \citenamefont
  {Vishwanath}}]{Yao_dtc:2016wp}%
  \BibitemOpen
  \bibfield  {author} {\bibinfo {author} {\bibfnamefont {N.~Y.}\ \bibnamefont
  {Yao}}, \bibinfo {author} {\bibfnamefont {A.~C.}\ \bibnamefont {Potter}},
  \bibinfo {author} {\bibfnamefont {I.-D.}\ \bibnamefont {Potirniche}}, \ and\
  \bibinfo {author} {\bibfnamefont {A.}~\bibnamefont {Vishwanath}},\
  }\href@noop {} {\bibfield  {journal} {\bibinfo  {journal} {Physical Review
  Letters}\ }\textbf {\bibinfo {volume} {118}},\ \bibinfo {pages} {030401}
  (\bibinfo {year} {2017})}\BibitemShut {NoStop}%
\bibitem [{\citenamefont {Zhang}\ \emph {et~al.}(2017)\citenamefont {Zhang},
  \citenamefont {Hess}, \citenamefont {Kyprianidis}, \citenamefont {Becker},
  \citenamefont {Lee}, \citenamefont {Smith}, \citenamefont {Pagano},
  \citenamefont {Potirniche}, \citenamefont {Potter}, \citenamefont
  {Vishwanath}, \citenamefont {Yao},\ and\ \citenamefont
  {Monroe}}]{Zhang:2016uw}%
  \BibitemOpen
  \bibfield  {author} {\bibinfo {author} {\bibfnamefont {J.}~\bibnamefont
  {Zhang}}, \bibinfo {author} {\bibfnamefont {P.~W.}\ \bibnamefont {Hess}},
  \bibinfo {author} {\bibfnamefont {A.}~\bibnamefont {Kyprianidis}}, \bibinfo
  {author} {\bibfnamefont {P.}~\bibnamefont {Becker}}, \bibinfo {author}
  {\bibfnamefont {A.}~\bibnamefont {Lee}}, \bibinfo {author} {\bibfnamefont
  {J.}~\bibnamefont {Smith}}, \bibinfo {author} {\bibfnamefont
  {G.}~\bibnamefont {Pagano}}, \bibinfo {author} {\bibfnamefont {I.~D.}\
  \bibnamefont {Potirniche}}, \bibinfo {author} {\bibfnamefont {A.~C.}\
  \bibnamefont {Potter}}, \bibinfo {author} {\bibfnamefont {A.}~\bibnamefont
  {Vishwanath}}, \bibinfo {author} {\bibfnamefont {N.~Y.}\ \bibnamefont {Yao}},
  \ and\ \bibinfo {author} {\bibfnamefont {C.}~\bibnamefont {Monroe}},\
  }\href@noop {} {\bibfield  {journal} {\bibinfo  {journal} {Nature}\ }\textbf
  {\bibinfo {volume} {543}},\ \bibinfo {pages} {217} (\bibinfo {year}
  {2017})}\BibitemShut {NoStop}%
\bibitem [{\citenamefont {Choi}\ \emph
  {et~al.}(2017{\natexlab{a}})\citenamefont {Choi}, \citenamefont {Choi},
  \citenamefont {Landig}, \citenamefont {Kucsko}, \citenamefont {Zhou},
  \citenamefont {Isoya}, \citenamefont {Jelezko}, \citenamefont {Onoda},
  \citenamefont {Sumiya}, \citenamefont {Khemani}, \citenamefont {von
  Keyserlingk}, \citenamefont {Yao}, \citenamefont {Demler},\ and\
  \citenamefont {Lukin}}]{Choi:2016wn}%
  \BibitemOpen
  \bibfield  {author} {\bibinfo {author} {\bibfnamefont {S.}~\bibnamefont
  {Choi}}, \bibinfo {author} {\bibfnamefont {J.}~\bibnamefont {Choi}}, \bibinfo
  {author} {\bibfnamefont {R.}~\bibnamefont {Landig}}, \bibinfo {author}
  {\bibfnamefont {G.}~\bibnamefont {Kucsko}}, \bibinfo {author} {\bibfnamefont
  {H.}~\bibnamefont {Zhou}}, \bibinfo {author} {\bibfnamefont {J.}~\bibnamefont
  {Isoya}}, \bibinfo {author} {\bibfnamefont {F.}~\bibnamefont {Jelezko}},
  \bibinfo {author} {\bibfnamefont {S.}~\bibnamefont {Onoda}}, \bibinfo
  {author} {\bibfnamefont {H.}~\bibnamefont {Sumiya}}, \bibinfo {author}
  {\bibfnamefont {V.}~\bibnamefont {Khemani}}, \bibinfo {author} {\bibfnamefont
  {C.}~\bibnamefont {von Keyserlingk}}, \bibinfo {author} {\bibfnamefont
  {N.~Y.}\ \bibnamefont {Yao}}, \bibinfo {author} {\bibfnamefont {E.~A.}\
  \bibnamefont {Demler}}, \ and\ \bibinfo {author} {\bibfnamefont {M.~D.}\
  \bibnamefont {Lukin}},\ }\href@noop {} {\bibfield  {journal} {\bibinfo
  {journal} {Nature}\ }\textbf {\bibinfo {volume} {543}},\ \bibinfo {pages}
  {221} (\bibinfo {year} {2017}{\natexlab{a}})}\BibitemShut {NoStop}%
\bibitem [{\citenamefont {Ho}\ \emph {et~al.}(2017)\citenamefont {Ho},
  \citenamefont {Choi}, \citenamefont {Lukin},\ and\ \citenamefont
  {Abanin}}]{Ho:2017ea}%
  \BibitemOpen
  \bibfield  {author} {\bibinfo {author} {\bibfnamefont {W.~W.}\ \bibnamefont
  {Ho}}, \bibinfo {author} {\bibfnamefont {S.}~\bibnamefont {Choi}}, \bibinfo
  {author} {\bibfnamefont {M.~D.}\ \bibnamefont {Lukin}}, \ and\ \bibinfo
  {author} {\bibfnamefont {D.~A.}\ \bibnamefont {Abanin}},\ }\href@noop {}
  {\bibfield  {journal} {\bibinfo  {journal} {Physical Review Letters}\
  }\textbf {\bibinfo {volume} {119}},\ \bibinfo {pages} {010602} (\bibinfo
  {year} {2017})}\BibitemShut {NoStop}%
\bibitem [{\citenamefont {Else}\ \emph {et~al.}(2017)\citenamefont {Else},
  \citenamefont {Bauer},\ and\ \citenamefont {Nayak}}]{Else:2017kg}%
  \BibitemOpen
  \bibfield  {author} {\bibinfo {author} {\bibfnamefont {D.~V.}\ \bibnamefont
  {Else}}, \bibinfo {author} {\bibfnamefont {B.}~\bibnamefont {Bauer}}, \ and\
  \bibinfo {author} {\bibfnamefont {C.}~\bibnamefont {Nayak}},\ }\href@noop {}
  {\bibfield  {journal} {\bibinfo  {journal} {Physical Review X}\ }\textbf
  {\bibinfo {volume} {7}},\ \bibinfo {pages} {011026} (\bibinfo {year}
  {2017})}\BibitemShut {NoStop}%
\bibitem [{\citenamefont {Allred}\ \emph {et~al.}(2002)\citenamefont {Allred},
  \citenamefont {Lyman}, \citenamefont {Kornack},\ and\ \citenamefont
  {Romalis}}]{Allred:2002bj}%
  \BibitemOpen
  \bibfield  {author} {\bibinfo {author} {\bibfnamefont {J.~C.}\ \bibnamefont
  {Allred}}, \bibinfo {author} {\bibfnamefont {R.~N.}\ \bibnamefont {Lyman}},
  \bibinfo {author} {\bibfnamefont {T.~W.}\ \bibnamefont {Kornack}}, \ and\
  \bibinfo {author} {\bibfnamefont {M.~V.}\ \bibnamefont {Romalis}},\
  }\href@noop {} {\bibfield  {journal} {\bibinfo  {journal} {Physical Review
  Letters}\ }\textbf {\bibinfo {volume} {89}},\ \bibinfo {pages} {130801}
  (\bibinfo {year} {2002})}\BibitemShut {NoStop}%
\bibitem [{\citenamefont {Deutsch}\ \emph {et~al.}(2010)\citenamefont
  {Deutsch}, \citenamefont {Ramirez-Martinez}, \citenamefont {Lacro{\^u}te},
  \citenamefont {Reinhard}, \citenamefont {Schneider}, \citenamefont {Fuchs},
  \citenamefont {Pi{\'e}chon}, \citenamefont {Lalo{\"e}}, \citenamefont
  {Reichel},\ and\ \citenamefont {Rosenbusch}}]{Deutsch:2010ky}%
  \BibitemOpen
  \bibfield  {author} {\bibinfo {author} {\bibfnamefont {C.}~\bibnamefont
  {Deutsch}}, \bibinfo {author} {\bibfnamefont {F.}~\bibnamefont
  {Ramirez-Martinez}}, \bibinfo {author} {\bibfnamefont {C.}~\bibnamefont
  {Lacro{\^u}te}}, \bibinfo {author} {\bibfnamefont {F.}~\bibnamefont
  {Reinhard}}, \bibinfo {author} {\bibfnamefont {T.}~\bibnamefont {Schneider}},
  \bibinfo {author} {\bibfnamefont {J.~N.}\ \bibnamefont {Fuchs}}, \bibinfo
  {author} {\bibfnamefont {F.}~\bibnamefont {Pi{\'e}chon}}, \bibinfo {author}
  {\bibfnamefont {F.}~\bibnamefont {Lalo{\"e}}}, \bibinfo {author}
  {\bibfnamefont {J.}~\bibnamefont {Reichel}}, \ and\ \bibinfo {author}
  {\bibfnamefont {P.}~\bibnamefont {Rosenbusch}},\ }\href@noop {} {\bibfield
  {journal} {\bibinfo  {journal} {Physical Review Letters}\ }\textbf {\bibinfo
  {volume} {105}},\ \bibinfo {pages} {020401} (\bibinfo {year}
  {2010})}\BibitemShut {NoStop}%
\bibitem [{Note1()}]{Note1}%
  \BibitemOpen
  \bibinfo {note} {We note that our approach does not change the SQL
  sensitivity scaling. However, if the external noise which limits $T_2$
  exhibits spatial correlations, it is well known that one can achieve an
  enhanced scaling with $N$ \cite {Paola:2017RMP_sensing}.}\BibitemShut {Stop}%
\bibitem [{\citenamefont {Zanardi}\ \emph {et~al.}(2008)\citenamefont
  {Zanardi}, \citenamefont {Paris},\ and\ \citenamefont
  {Venuti}}]{Zanardi:2008ih}%
  \BibitemOpen
  \bibfield  {author} {\bibinfo {author} {\bibfnamefont {P.}~\bibnamefont
  {Zanardi}}, \bibinfo {author} {\bibfnamefont {M.~G.~A.}\ \bibnamefont
  {Paris}}, \ and\ \bibinfo {author} {\bibfnamefont {L.~C.}\ \bibnamefont
  {Venuti}},\ }\href@noop {} {\bibfield  {journal} {\bibinfo  {journal}
  {Physical Review A}\ }\textbf {\bibinfo {volume} {78}},\ \bibinfo {pages}
  {042105} (\bibinfo {year} {2008})}\BibitemShut {NoStop}%
\bibitem [{Mac(2016)}]{Macieszczak:2016gw}%
  \BibitemOpen
  \href@noop {} {\ \textbf {\bibinfo {volume} {93}},\ \bibinfo {pages} {022103}
  (\bibinfo {year} {2016})}\BibitemShut {NoStop}%
\bibitem [{\citenamefont {Skotiniotis}\ \emph {et~al.}(2015)\citenamefont
  {Skotiniotis}, \citenamefont {Sekatski},\ and\ \citenamefont
  {D{\"u}r}}]{Skotiniotis:2015go}%
  \BibitemOpen
  \bibfield  {author} {\bibinfo {author} {\bibfnamefont {M.}~\bibnamefont
  {Skotiniotis}}, \bibinfo {author} {\bibfnamefont {P.}~\bibnamefont
  {Sekatski}}, \ and\ \bibinfo {author} {\bibfnamefont {W.}~\bibnamefont
  {D{\"u}r}},\ }\href@noop {} {\bibfield  {journal} {\bibinfo  {journal} {New
  Journal of Physics}\ }\textbf {\bibinfo {volume} {17}},\ \bibinfo {pages}
  {073032} (\bibinfo {year} {2015})}\BibitemShut {NoStop}%
\bibitem [{\citenamefont {Fr{\'e}rot}\ and\ \citenamefont
  {Roscilde}(2017)}]{Frerot:2017uh}%
  \BibitemOpen
  \bibfield  {author} {\bibinfo {author} {\bibfnamefont {I.}~\bibnamefont
  {Fr{\'e}rot}}\ and\ \bibinfo {author} {\bibfnamefont {T.}~\bibnamefont
  {Roscilde}},\ }\href@noop {} {\  (\bibinfo {year} {2017})}\BibitemShut
  {NoStop}%
\bibitem [{\citenamefont {Strobel}\ \emph {et~al.}(2014)\citenamefont
  {Strobel}, \citenamefont {Muessel}, \citenamefont {Linnemann}, \citenamefont
  {Zibold}, \citenamefont {Hume}, \citenamefont {Pezz{\`e}}, \citenamefont
  {Smerzi},\ and\ \citenamefont {Oberthaler}}]{Strobel:2014eg}%
  \BibitemOpen
  \bibfield  {author} {\bibinfo {author} {\bibfnamefont {H.}~\bibnamefont
  {Strobel}}, \bibinfo {author} {\bibfnamefont {W.}~\bibnamefont {Muessel}},
  \bibinfo {author} {\bibfnamefont {D.}~\bibnamefont {Linnemann}}, \bibinfo
  {author} {\bibfnamefont {T.}~\bibnamefont {Zibold}}, \bibinfo {author}
  {\bibfnamefont {D.~B.}\ \bibnamefont {Hume}}, \bibinfo {author}
  {\bibfnamefont {L.}~\bibnamefont {Pezz{\`e}}}, \bibinfo {author}
  {\bibfnamefont {A.}~\bibnamefont {Smerzi}}, \ and\ \bibinfo {author}
  {\bibfnamefont {M.~K.}\ \bibnamefont {Oberthaler}},\ }\href@noop {}
  {\bibfield  {journal} {\bibinfo  {journal} {Science}\ }\textbf {\bibinfo
  {volume} {345}},\ \bibinfo {pages} {424} (\bibinfo {year}
  {2014})}\BibitemShut {NoStop}%
\bibitem [{\citenamefont {Hosten}\ \emph {et~al.}(2016)\citenamefont {Hosten},
  \citenamefont {Engelsen}, \citenamefont {Krishnakumar},\ and\ \citenamefont
  {Kasevich}}]{Hosten:2016dj}%
  \BibitemOpen
  \bibfield  {author} {\bibinfo {author} {\bibfnamefont {O.}~\bibnamefont
  {Hosten}}, \bibinfo {author} {\bibfnamefont {N.~J.}\ \bibnamefont
  {Engelsen}}, \bibinfo {author} {\bibfnamefont {R.}~\bibnamefont
  {Krishnakumar}}, \ and\ \bibinfo {author} {\bibfnamefont {M.~A.}\
  \bibnamefont {Kasevich}},\ }\href@noop {} {\bibfield  {journal} {\bibinfo
  {journal} {Nature}\ }\textbf {\bibinfo {volume} {529}},\ \bibinfo {pages}
  {505} (\bibinfo {year} {2016})}\BibitemShut {NoStop}%
\bibitem [{\citenamefont {Bohnet}\ \emph {et~al.}(2016)\citenamefont {Bohnet},
  \citenamefont {Sawyer}, \citenamefont {Britton}, \citenamefont {Wall},
  \citenamefont {Rey}, \citenamefont {Foss-Feig},\ and\ \citenamefont
  {Bollinger}}]{Bohnet:2016ej}%
  \BibitemOpen
  \bibfield  {author} {\bibinfo {author} {\bibfnamefont {J.~G.}\ \bibnamefont
  {Bohnet}}, \bibinfo {author} {\bibfnamefont {B.~C.}\ \bibnamefont {Sawyer}},
  \bibinfo {author} {\bibfnamefont {J.~W.}\ \bibnamefont {Britton}}, \bibinfo
  {author} {\bibfnamefont {M.~L.}\ \bibnamefont {Wall}}, \bibinfo {author}
  {\bibfnamefont {A.~M.}\ \bibnamefont {Rey}}, \bibinfo {author} {\bibfnamefont
  {M.}~\bibnamefont {Foss-Feig}}, \ and\ \bibinfo {author} {\bibfnamefont
  {J.~J.}\ \bibnamefont {Bollinger}},\ }\href@noop {} {\bibfield  {journal}
  {\bibinfo  {journal} {Science}\ }\textbf {\bibinfo {volume} {352}},\ \bibinfo
  {pages} {1297} (\bibinfo {year} {2016})}\BibitemShut {NoStop}%
\bibitem [{\citenamefont {Aasi}\ \emph {et~al.}(2013)\citenamefont {Aasi} \emph
  {et~al.}}]{Aasi:2013jb}%
  \BibitemOpen
  \bibfield  {author} {\bibinfo {author} {\bibfnamefont {J.}~\bibnamefont
  {Aasi}} \emph {et~al.},\ }\href@noop {} {\bibfield  {journal} {\bibinfo
  {journal} {Nature Photonics}\ }\textbf {\bibinfo {volume} {7}},\ \bibinfo
  {pages} {613} (\bibinfo {year} {2013})}\BibitemShut {NoStop}%
\bibitem [{\citenamefont {Hahn}(1950)}]{Hahn_echo:1950ge}%
  \BibitemOpen
  \bibfield  {author} {\bibinfo {author} {\bibfnamefont {E.~L.}\ \bibnamefont
  {Hahn}},\ }\href@noop {} {\bibfield  {journal} {\bibinfo  {journal} {Physical
  Review}\ }\textbf {\bibinfo {volume} {80}},\ \bibinfo {pages} {580} (\bibinfo
  {year} {1950})}\BibitemShut {NoStop}%
\bibitem [{\citenamefont {de~Lange}\ \emph {et~al.}(2010)\citenamefont
  {de~Lange}, \citenamefont {Wang}, \citenamefont {Rist{\`e}}, \citenamefont
  {Dobrovitski},\ and\ \citenamefont {Hanson}}]{deLange:2010ga}%
  \BibitemOpen
  \bibfield  {author} {\bibinfo {author} {\bibfnamefont {G.}~\bibnamefont
  {de~Lange}}, \bibinfo {author} {\bibfnamefont {Z.~H.}\ \bibnamefont {Wang}},
  \bibinfo {author} {\bibfnamefont {D.}~\bibnamefont {Rist{\`e}}}, \bibinfo
  {author} {\bibfnamefont {V.~V.}\ \bibnamefont {Dobrovitski}}, \ and\ \bibinfo
  {author} {\bibfnamefont {R.}~\bibnamefont {Hanson}},\ }\href@noop {}
  {\bibfield  {journal} {\bibinfo  {journal} {Science}\ }\textbf {\bibinfo
  {volume} {330}},\ \bibinfo {pages} {60} (\bibinfo {year} {2010})}\BibitemShut
  {NoStop}%
\bibitem [{\citenamefont {Kuo}\ and\ \citenamefont
  {Lidar}(2011)}]{QDDproofLidar:2011aa}%
  \BibitemOpen
  \bibfield  {author} {\bibinfo {author} {\bibfnamefont {W.-J.}\ \bibnamefont
  {Kuo}}\ and\ \bibinfo {author} {\bibfnamefont {D.~A.}\ \bibnamefont
  {Lidar}},\ }\href@noop {} {\bibfield  {journal} {\bibinfo  {journal}
  {Physical Review A}\ }\textbf {\bibinfo {volume} {84}},\ \bibinfo {pages}
  {042329} (\bibinfo {year} {2011})}\BibitemShut {NoStop}%
\bibitem [{\citenamefont {Jiang}\ and\ \citenamefont
  {Imambekov}(2011)}]{UDDproofJiang:2011aa}%
  \BibitemOpen
  \bibfield  {author} {\bibinfo {author} {\bibfnamefont {L.}~\bibnamefont
  {Jiang}}\ and\ \bibinfo {author} {\bibfnamefont {A.}~\bibnamefont
  {Imambekov}},\ }\href@noop {} {\bibfield  {journal} {\bibinfo  {journal}
  {Physical Review A}\ }\textbf {\bibinfo {volume} {84}},\ \bibinfo {pages}
  {060302} (\bibinfo {year} {2011})}\BibitemShut {NoStop}%
\bibitem [{\citenamefont {Choi}\ \emph
  {et~al.}(2017{\natexlab{b}})\citenamefont {Choi}, \citenamefont {Yao},\ and\
  \citenamefont {Lukin}}]{Choi:2017dynamical_engineering}%
  \BibitemOpen
  \bibfield  {author} {\bibinfo {author} {\bibfnamefont {S.}~\bibnamefont
  {Choi}}, \bibinfo {author} {\bibfnamefont {N.~Y.}\ \bibnamefont {Yao}}, \
  and\ \bibinfo {author} {\bibfnamefont {M.~D.}\ \bibnamefont {Lukin}},\
  }\href@noop {} {\bibfield  {journal} {\bibinfo  {journal} {Physical Review
  Letters}\ }\textbf {\bibinfo {volume} {119}},\ \bibinfo {pages} {183603}
  (\bibinfo {year} {2017}{\natexlab{b}})}\BibitemShut {NoStop}%
\bibitem [{\citenamefont {Maurer}\ \emph {et~al.}(2012)\citenamefont {Maurer},
  \citenamefont {Kucsko}, \citenamefont {Latta}, \citenamefont {Jiang},
  \citenamefont {Yao}, \citenamefont {Bennett}, \citenamefont {Pastawski},
  \citenamefont {Hunger}, \citenamefont {Chisholm}, \citenamefont {Markham},
  \citenamefont {Twitchen}, \citenamefont {Cirac},\ and\ \citenamefont
  {Lukin}}]{NV_memory:2012Maurer}%
  \BibitemOpen
  \bibfield  {author} {\bibinfo {author} {\bibfnamefont {P.~C.}\ \bibnamefont
  {Maurer}}, \bibinfo {author} {\bibfnamefont {G.}~\bibnamefont {Kucsko}},
  \bibinfo {author} {\bibfnamefont {C.}~\bibnamefont {Latta}}, \bibinfo
  {author} {\bibfnamefont {L.}~\bibnamefont {Jiang}}, \bibinfo {author}
  {\bibfnamefont {N.~Y.}\ \bibnamefont {Yao}}, \bibinfo {author} {\bibfnamefont
  {S.~D.}\ \bibnamefont {Bennett}}, \bibinfo {author} {\bibfnamefont
  {F.}~\bibnamefont {Pastawski}}, \bibinfo {author} {\bibfnamefont
  {D.}~\bibnamefont {Hunger}}, \bibinfo {author} {\bibfnamefont
  {N.}~\bibnamefont {Chisholm}}, \bibinfo {author} {\bibfnamefont
  {M.}~\bibnamefont {Markham}}, \bibinfo {author} {\bibfnamefont {D.~J.}\
  \bibnamefont {Twitchen}}, \bibinfo {author} {\bibfnamefont {J.~I.}\
  \bibnamefont {Cirac}}, \ and\ \bibinfo {author} {\bibfnamefont {M.~D.}\
  \bibnamefont {Lukin}},\ }\href@noop {} {\bibfield  {journal} {\bibinfo
  {journal} {Science}\ }\textbf {\bibinfo {volume} {336}},\ \bibinfo {pages}
  {1283} (\bibinfo {year} {2012})}\BibitemShut {NoStop}%
\bibitem [{\citenamefont {Lovchinsky}\ \emph {et~al.}(2016)\citenamefont
  {Lovchinsky}, \citenamefont {Sushkov}, \citenamefont {Urbach}, \citenamefont
  {de~Leon}, \citenamefont {Choi}, \citenamefont {De~Greve}, \citenamefont
  {Evans}, \citenamefont {Gertner}, \citenamefont {Bersin}, \citenamefont
  {M{\"u}ller}, \citenamefont {McGuinness}, \citenamefont {Jelezko},
  \citenamefont {Walsworth}, \citenamefont {Park},\ and\ \citenamefont
  {Lukin}}]{Lovchinsky:2016dq}%
  \BibitemOpen
  \bibfield  {author} {\bibinfo {author} {\bibfnamefont {I.}~\bibnamefont
  {Lovchinsky}}, \bibinfo {author} {\bibfnamefont {A.~O.}\ \bibnamefont
  {Sushkov}}, \bibinfo {author} {\bibfnamefont {E.}~\bibnamefont {Urbach}},
  \bibinfo {author} {\bibfnamefont {N.~P.}\ \bibnamefont {de~Leon}}, \bibinfo
  {author} {\bibfnamefont {S.}~\bibnamefont {Choi}}, \bibinfo {author}
  {\bibfnamefont {K.}~\bibnamefont {De~Greve}}, \bibinfo {author}
  {\bibfnamefont {R.}~\bibnamefont {Evans}}, \bibinfo {author} {\bibfnamefont
  {R.}~\bibnamefont {Gertner}}, \bibinfo {author} {\bibfnamefont
  {E.}~\bibnamefont {Bersin}}, \bibinfo {author} {\bibfnamefont
  {C.}~\bibnamefont {M{\"u}ller}}, \bibinfo {author} {\bibfnamefont
  {L.}~\bibnamefont {McGuinness}}, \bibinfo {author} {\bibfnamefont
  {F.}~\bibnamefont {Jelezko}}, \bibinfo {author} {\bibfnamefont {R.~L.}\
  \bibnamefont {Walsworth}}, \bibinfo {author} {\bibfnamefont {H.}~\bibnamefont
  {Park}}, \ and\ \bibinfo {author} {\bibfnamefont {M.~D.}\ \bibnamefont
  {Lukin}},\ }\href@noop {} {\bibfield  {journal} {\bibinfo  {journal}
  {Science}\ }\textbf {\bibinfo {volume} {351}},\ \bibinfo {pages} {836}
  (\bibinfo {year} {2016})}\BibitemShut {NoStop}%
\bibitem [{\citenamefont {Lovchinsky}\ \emph {et~al.}(2017)\citenamefont
  {Lovchinsky}, \citenamefont {Sanchez-Yamagishi}, \citenamefont {Urbach},
  \citenamefont {Choi}, \citenamefont {Fang}, \citenamefont {Andersen},
  \citenamefont {Watanabe}, \citenamefont {Taniguchi}, \citenamefont
  {Bylinskii}, \citenamefont {Kaxiras}, \citenamefont {Kim}, \citenamefont
  {Park},\ and\ \citenamefont {Lukin}}]{Lovchinsky:2017gh}%
  \BibitemOpen
  \bibfield  {author} {\bibinfo {author} {\bibfnamefont {I.}~\bibnamefont
  {Lovchinsky}}, \bibinfo {author} {\bibfnamefont {J.~D.}\ \bibnamefont
  {Sanchez-Yamagishi}}, \bibinfo {author} {\bibfnamefont {E.~K.}\ \bibnamefont
  {Urbach}}, \bibinfo {author} {\bibfnamefont {S.}~\bibnamefont {Choi}},
  \bibinfo {author} {\bibfnamefont {S.}~\bibnamefont {Fang}}, \bibinfo {author}
  {\bibfnamefont {T.~I.}\ \bibnamefont {Andersen}}, \bibinfo {author}
  {\bibfnamefont {K.}~\bibnamefont {Watanabe}}, \bibinfo {author}
  {\bibfnamefont {T.}~\bibnamefont {Taniguchi}}, \bibinfo {author}
  {\bibfnamefont {A.}~\bibnamefont {Bylinskii}}, \bibinfo {author}
  {\bibfnamefont {E.}~\bibnamefont {Kaxiras}}, \bibinfo {author} {\bibfnamefont
  {P.}~\bibnamefont {Kim}}, \bibinfo {author} {\bibfnamefont {H.}~\bibnamefont
  {Park}}, \ and\ \bibinfo {author} {\bibfnamefont {M.~D.}\ \bibnamefont
  {Lukin}},\ }\href@noop {} {\bibfield  {journal} {\bibinfo  {journal}
  {Science}\ }\textbf {\bibinfo {volume} {355}},\ \bibinfo {pages} {503}
  (\bibinfo {year} {2017})}\BibitemShut {NoStop}%
\bibitem [{Note2()}]{Note2}%
  \BibitemOpen
  \bibinfo {note} {More specifically we require $|\omega _0 - 2 \omega _s| \gg
  \Omega , J$.}\BibitemShut {Stop}%
\bibitem [{Note3()}]{Note3}%
  \BibitemOpen
  \bibinfo {note} {Note that the parity operator $P$ conicides with the unitary
  that globally rotates the spin ensemble by $\pi $ up to an unimportant
  complex phase.}\BibitemShut {Stop}%
\bibitem [{\citenamefont {Slichter}(2013)}]{Slichter:2013wo}%
  \BibitemOpen
  \bibfield  {author} {\bibinfo {author} {\bibfnamefont {C.~P.}\ \bibnamefont
  {Slichter}},\ }\href@noop {} {\emph {\bibinfo {title} {{Principles of
  magnetic resonance}}}},\ Vol.~\bibinfo {volume} {1}\ (\bibinfo  {publisher}
  {Springer Science {\&} Business Media},\ \bibinfo {year} {2013})\BibitemShut
  {NoStop}%
\bibitem [{\citenamefont {Kuwahara}\ \emph {et~al.}(2016)\citenamefont
  {Kuwahara}, \citenamefont {Mori},\ and\ \citenamefont
  {Saito}}]{Kuwahara:2016dh}%
  \BibitemOpen
  \bibfield  {author} {\bibinfo {author} {\bibfnamefont {T.}~\bibnamefont
  {Kuwahara}}, \bibinfo {author} {\bibfnamefont {T.}~\bibnamefont {Mori}}, \
  and\ \bibinfo {author} {\bibfnamefont {K.}~\bibnamefont {Saito}},\
  }\href@noop {} {\bibfield  {journal} {\bibinfo  {journal} {Annals of
  Physics}\ }\textbf {\bibinfo {volume} {367}},\ \bibinfo {pages} {96}
  (\bibinfo {year} {2016})}\BibitemShut {NoStop}%
\bibitem [{\citenamefont {Abanin}\ \emph {et~al.}(2017)\citenamefont {Abanin},
  \citenamefont {De~Roeck}, \citenamefont {Ho},\ and\ \citenamefont
  {Huveneers}}]{Abanin:2017hp}%
  \BibitemOpen
  \bibfield  {author} {\bibinfo {author} {\bibfnamefont {D.}~\bibnamefont
  {Abanin}}, \bibinfo {author} {\bibfnamefont {W.}~\bibnamefont {De~Roeck}},
  \bibinfo {author} {\bibfnamefont {W.~W.}\ \bibnamefont {Ho}}, \ and\ \bibinfo
  {author} {\bibfnamefont {F.}~\bibnamefont {Huveneers}},\ }\href@noop {}
  {\bibfield  {journal} {\bibinfo  {journal} {Communications in Mathematical
  Physics}\ }\textbf {\bibinfo {volume} {354}},\ \bibinfo {pages} {809}
  (\bibinfo {year} {2017})}\BibitemShut {NoStop}%
\bibitem [{Note4()}]{Note4}%
  \BibitemOpen
  \bibinfo {note} {This description is valid up to an exponentially long time
  $\sim \protect \qopname \relax o{exp}{[\omega _0/\protect \textrm
  {max}(\Omega ,J)]}$, beyond which the system absorbs energy from the periodic
  driving and heats up to infinite temperature~\cite
  {Abanin:2015bc,Mori:2016wb,Kuwahara:2016dh,Abanin:2017hp}.}\BibitemShut
  {Stop}%
\bibitem [{\citenamefont {Zurek}\ \emph {et~al.}(2005)\citenamefont {Zurek},
  \citenamefont {Dorner},\ and\ \citenamefont {Zoller}}]{Zurek:2005cu}%
  \BibitemOpen
  \bibfield  {author} {\bibinfo {author} {\bibfnamefont {W.~H.}\ \bibnamefont
  {Zurek}}, \bibinfo {author} {\bibfnamefont {U.}~\bibnamefont {Dorner}}, \
  and\ \bibinfo {author} {\bibfnamefont {P.}~\bibnamefont {Zoller}},\
  }\href@noop {} {\bibfield  {journal} {\bibinfo  {journal} {Physical Review
  Letters}\ }\textbf {\bibinfo {volume} {95}},\ \bibinfo {pages} {105701}
  (\bibinfo {year} {2005})}\BibitemShut {NoStop}%
\bibitem [{\citenamefont {Fisher}\ \emph {et~al.}(1972)\citenamefont {Fisher},
  \citenamefont {Ma},\ and\ \citenamefont {Nickel}}]{Fisher:1972gh}%
  \BibitemOpen
  \bibfield  {author} {\bibinfo {author} {\bibfnamefont {M.~E.}\ \bibnamefont
  {Fisher}}, \bibinfo {author} {\bibfnamefont {S.-k.}\ \bibnamefont {Ma}}, \
  and\ \bibinfo {author} {\bibfnamefont {B.~G.}\ \bibnamefont {Nickel}},\
  }\href@noop {} {\bibfield  {journal} {\bibinfo  {journal} {Physical Review
  Letters}\ }\textbf {\bibinfo {volume} {29}},\ \bibinfo {pages} {917}
  (\bibinfo {year} {1972})}\BibitemShut {NoStop}%
\bibitem [{\citenamefont {Dutta}\ and\ \citenamefont
  {Bhattacharjee}(2001)}]{Dutta:2001gu}%
  \BibitemOpen
  \bibfield  {author} {\bibinfo {author} {\bibfnamefont {A.}~\bibnamefont
  {Dutta}}\ and\ \bibinfo {author} {\bibfnamefont {J.~K.}\ \bibnamefont
  {Bhattacharjee}},\ }\href@noop {} {\bibfield  {journal} {\bibinfo  {journal}
  {Physical Review B}\ }\textbf {\bibinfo {volume} {64}},\ \bibinfo {pages}
  {184106} (\bibinfo {year} {2001})}\BibitemShut {NoStop}%
\bibitem [{\citenamefont {Knap}\ \emph {et~al.}(2013)\citenamefont {Knap},
  \citenamefont {Kantian}, \citenamefont {Giamarchi}, \citenamefont {Bloch},
  \citenamefont {Lukin},\ and\ \citenamefont {Demler}}]{Knap:2013jy}%
  \BibitemOpen
  \bibfield  {author} {\bibinfo {author} {\bibfnamefont {M.}~\bibnamefont
  {Knap}}, \bibinfo {author} {\bibfnamefont {A.}~\bibnamefont {Kantian}},
  \bibinfo {author} {\bibfnamefont {T.}~\bibnamefont {Giamarchi}}, \bibinfo
  {author} {\bibfnamefont {I.}~\bibnamefont {Bloch}}, \bibinfo {author}
  {\bibfnamefont {M.~D.}\ \bibnamefont {Lukin}}, \ and\ \bibinfo {author}
  {\bibfnamefont {E.}~\bibnamefont {Demler}},\ }\href@noop {} {\bibfield
  {journal} {\bibinfo  {journal} {Physical Review Letters}\ }\textbf {\bibinfo
  {volume} {111}},\ \bibinfo {pages} {147205} (\bibinfo {year}
  {2013})}\BibitemShut {NoStop}%
\bibitem [{\citenamefont {Fey}\ and\ \citenamefont
  {Schmidt}(2016)}]{Fey:2016er}%
  \BibitemOpen
  \bibfield  {author} {\bibinfo {author} {\bibfnamefont {S.}~\bibnamefont
  {Fey}}\ and\ \bibinfo {author} {\bibfnamefont {K.~P.}\ \bibnamefont
  {Schmidt}},\ }\href@noop {} {\bibfield  {journal} {\bibinfo  {journal}
  {Physical Review B}\ }\textbf {\bibinfo {volume} {94}},\ \bibinfo {pages}
  {075156} (\bibinfo {year} {2016})}\BibitemShut {NoStop}%
\bibitem [{\citenamefont {Maghrebi}\ \emph {et~al.}(2016)\citenamefont
  {Maghrebi}, \citenamefont {Gong}, \citenamefont {Foss-Feig},\ and\
  \citenamefont {Gorshkov}}]{Maghrebi:2016bp}%
  \BibitemOpen
  \bibfield  {author} {\bibinfo {author} {\bibfnamefont {M.~F.}\ \bibnamefont
  {Maghrebi}}, \bibinfo {author} {\bibfnamefont {Z.-X.}\ \bibnamefont {Gong}},
  \bibinfo {author} {\bibfnamefont {M.}~\bibnamefont {Foss-Feig}}, \ and\
  \bibinfo {author} {\bibfnamefont {A.~V.}\ \bibnamefont {Gorshkov}},\
  }\href@noop {} {\bibfield  {journal} {\bibinfo  {journal} {Physical Review
  B}\ }\textbf {\bibinfo {volume} {93}},\ \bibinfo {pages} {125128} (\bibinfo
  {year} {2016})}\BibitemShut {NoStop}%
\bibitem [{\citenamefont {Elliott}\ and\ \citenamefont
  {Wood}(1971)}]{Elliott:1971hr}%
  \BibitemOpen
  \bibfield  {author} {\bibinfo {author} {\bibfnamefont {R.~J.}\ \bibnamefont
  {Elliott}}\ and\ \bibinfo {author} {\bibfnamefont {C.}~\bibnamefont {Wood}},\
  }\href@noop {} {\bibfield  {journal} {\bibinfo  {journal} {Journal of Physics
  C: Solid State Physics}\ }\textbf {\bibinfo {volume} {4}},\ \bibinfo {pages}
  {2359} (\bibinfo {year} {1971})}\BibitemShut {NoStop}%
\bibitem [{\citenamefont {Pfeuty}\ and\ \citenamefont
  {Elliott}(1971)}]{Pfeuty:1971ie}%
  \BibitemOpen
  \bibfield  {author} {\bibinfo {author} {\bibfnamefont {P.}~\bibnamefont
  {Pfeuty}}\ and\ \bibinfo {author} {\bibfnamefont {R.~J.}\ \bibnamefont
  {Elliott}},\ }\href@noop {} {\bibfield  {journal} {\bibinfo  {journal}
  {Journal of Physics C: Solid State Physics}\ }\textbf {\bibinfo {volume}
  {4}},\ \bibinfo {pages} {2370} (\bibinfo {year} {1971})}\BibitemShut
  {NoStop}%
\bibitem [{\citenamefont {Friedman}(1978)}]{Friedman:1978jb}%
  \BibitemOpen
  \bibfield  {author} {\bibinfo {author} {\bibfnamefont {Z.}~\bibnamefont
  {Friedman}},\ }\href@noop {} {\bibfield  {journal} {\bibinfo  {journal}
  {Physical Review B}\ }\textbf {\bibinfo {volume} {17}},\ \bibinfo {pages}
  {1429} (\bibinfo {year} {1978})}\BibitemShut {NoStop}%
\bibitem [{Note5()}]{Note5}%
  \BibitemOpen
  \bibinfo {note} {Interestingly, the effect of disorder can be favorable
  during the measurement stage; while domain wall excitations can be mobile in
  the absence of disorder (in 1D), relatively weak disorder in $J$ may localize
  the excitations, allowing stable accumulation of phase information over long
  times. This effect is particularly relevant when the localization lengths
  during the measurement step is much shorter than that at the critical point,
  which is often satisfied in realistic systems, where the dominant source of
  disorder arises from random positioning of spins (disorder in $J$)~\cite
  {supp_info}.}\BibitemShut {Stop}%
\bibitem [{sup()}]{supp_info}%
  \BibitemOpen
  \href@noop {} {}\bibinfo {note} {See Supplementary Materials for detailed
  information}\BibitemShut {NoStop}%
\bibitem [{Note6()}]{Note6}%
  \BibitemOpen
  \bibinfo {note} {For this protocol, excitations should not be created during
  the initialization or read-out steps. This condition can be estimated from
  the Kibble-Zurek ``freezing point'' $\Delta \Omega \geq \Omega
  (JT_p)^{-1/(z\nu + 1)}$~\cite {Zurek:2005cu}.}\BibitemShut {Stop}%
\bibitem [{\citenamefont {Sachdev}(2011)}]{sachdev2011quantum}%
  \BibitemOpen
  \bibfield  {author} {\bibinfo {author} {\bibfnamefont {S.}~\bibnamefont
  {Sachdev}},\ }\href@noop {} {\emph {\bibinfo {title} {{Quantum phase
  transitions}}}}\ (\bibinfo  {publisher} {Cambridge University Press},\
  \bibinfo {address} {Cambridge New York},\ \bibinfo {year} {2011})\BibitemShut
  {NoStop}%
\bibitem [{\citenamefont {Hauke}\ \emph {et~al.}(2016)\citenamefont {Hauke},
  \citenamefont {Heyl}, \citenamefont {Tagliacozzo},\ and\ \citenamefont
  {Zoller}}]{Hauke:2016ht}%
  \BibitemOpen
  \bibfield  {author} {\bibinfo {author} {\bibfnamefont {P.}~\bibnamefont
  {Hauke}}, \bibinfo {author} {\bibfnamefont {M.}~\bibnamefont {Heyl}},
  \bibinfo {author} {\bibfnamefont {L.}~\bibnamefont {Tagliacozzo}}, \ and\
  \bibinfo {author} {\bibfnamefont {P.}~\bibnamefont {Zoller}},\ }\href@noop {}
  {\bibfield  {journal} {\bibinfo  {journal} {Nature Physics}\ }\textbf
  {\bibinfo {volume} {12}},\ \bibinfo {pages} {778} (\bibinfo {year}
  {2016})}\BibitemShut {NoStop}%
\bibitem [{\citenamefont {Bugrii}(2001)}]{Bugrii:2001bh}%
  \BibitemOpen
  \bibfield  {author} {\bibinfo {author} {\bibfnamefont {A.~I.}\ \bibnamefont
  {Bugrii}},\ }\href@noop {} {\bibfield  {journal} {\bibinfo  {journal}
  {Theoretical and Mathematical Physics}\ }\textbf {\bibinfo {volume} {127}},\
  \bibinfo {pages} {528} (\bibinfo {year} {2001})}\BibitemShut {NoStop}%
\bibitem [{\citenamefont {Fonseca}\ and\ \citenamefont
  {Zamolodchikov}(2003)}]{Fonseca:2003ev}%
  \BibitemOpen
  \bibfield  {author} {\bibinfo {author} {\bibfnamefont {P.}~\bibnamefont
  {Fonseca}}\ and\ \bibinfo {author} {\bibfnamefont {A.}~\bibnamefont
  {Zamolodchikov}},\ }\href@noop {} {\bibfield  {journal} {\bibinfo  {journal}
  {Journal of Statistical Physics}\ }\textbf {\bibinfo {volume} {110}},\
  \bibinfo {pages} {527} (\bibinfo {year} {2003})}\BibitemShut {NoStop}%
\bibitem [{\citenamefont {Essler}\ and\ \citenamefont
  {Konik}(2009)}]{Essler:2009dj}%
  \BibitemOpen
  \bibfield  {author} {\bibinfo {author} {\bibfnamefont {F.~H.~L.}\
  \bibnamefont {Essler}}\ and\ \bibinfo {author} {\bibfnamefont {R.~M.}\
  \bibnamefont {Konik}},\ }\href@noop {} {\bibfield  {journal} {\bibinfo
  {journal} {Journal of Statistical Mechanics: Theory and Experiment}\ }\textbf
  {\bibinfo {volume} {2009}},\ \bibinfo {pages} {P09018} (\bibinfo {year}
  {2009})}\BibitemShut {NoStop}%
\bibitem [{\citenamefont {Waugh}\ \emph {et~al.}(1968)\citenamefont {Waugh},
  \citenamefont {Huber},\ and\ \citenamefont {Haeberlen}}]{Waugh:1968im}%
  \BibitemOpen
  \bibfield  {author} {\bibinfo {author} {\bibfnamefont {J.~S.}\ \bibnamefont
  {Waugh}}, \bibinfo {author} {\bibfnamefont {L.~M.}\ \bibnamefont {Huber}}, \
  and\ \bibinfo {author} {\bibfnamefont {U.}~\bibnamefont {Haeberlen}},\
  }\href@noop {} {\bibfield  {journal} {\bibinfo  {journal} {Physical Review
  Letters}\ }\textbf {\bibinfo {volume} {20}},\ \bibinfo {pages} {180}
  (\bibinfo {year} {1968})}\BibitemShut {NoStop}%
\bibitem [{\citenamefont {Kolkowitz}\ \emph {et~al.}(2016)\citenamefont
  {Kolkowitz}, \citenamefont {Pikovski}, \citenamefont {Langellier},
  \citenamefont {Lukin}, \citenamefont {Walsworth},\ and\ \citenamefont
  {Ye}}]{Kolkowitz:2016gx}%
  \BibitemOpen
  \bibfield  {author} {\bibinfo {author} {\bibfnamefont {S.}~\bibnamefont
  {Kolkowitz}}, \bibinfo {author} {\bibfnamefont {I.}~\bibnamefont {Pikovski}},
  \bibinfo {author} {\bibfnamefont {N.}~\bibnamefont {Langellier}}, \bibinfo
  {author} {\bibfnamefont {M.~D.}\ \bibnamefont {Lukin}}, \bibinfo {author}
  {\bibfnamefont {R.~L.}\ \bibnamefont {Walsworth}}, \ and\ \bibinfo {author}
  {\bibfnamefont {J.}~\bibnamefont {Ye}},\ }\href@noop {} {\bibfield  {journal}
  {\bibinfo  {journal} {Physical Review D}\ }\textbf {\bibinfo {volume} {94}},\
  \bibinfo {pages} {124043} (\bibinfo {year} {2016})}\BibitemShut {NoStop}%
\end{thebibliography}%


\begin{thebibliography}{5}%
\makeatletter
\providecommand \@ifxundefined [1]{%
 \@ifx{#1\undefined}
}%
\providecommand \@ifnum [1]{%
 \ifnum #1\expandafter \@firstoftwo
 \else \expandafter \@secondoftwo
 \fi
}%
\providecommand \@ifx [1]{%
 \ifx #1\expandafter \@firstoftwo
 \else \expandafter \@secondoftwo
 \fi
}%
\providecommand \natexlab [1]{#1}%
\providecommand \enquote  [1]{``#1''}%
\providecommand \bibnamefont  [1]{#1}%
\providecommand \bibfnamefont [1]{#1}%
\providecommand \citenamefont [1]{#1}%
\providecommand \href@noop [0]{\@secondoftwo}%
\providecommand \href [0]{\begingroup \@sanitize@url \@href}%
\providecommand \@href[1]{\@@startlink{#1}\@@href}%
\providecommand \@@href[1]{\endgroup#1\@@endlink}%
\providecommand \@sanitize@url [0]{\catcode `\\12\catcode `\$12\catcode
  `\&12\catcode `\#12\catcode `\^12\catcode `\_12\catcode `\%12\relax}%
\providecommand \@@startlink[1]{}%
\providecommand \@@endlink[0]{}%
\providecommand \url  [0]{\begingroup\@sanitize@url \@url }%
\providecommand \@url [1]{\endgroup\@href {#1}{\urlprefix }}%
\providecommand \urlprefix  [0]{URL }%
\providecommand \Eprint [0]{\href }%
\providecommand \doibase [0]{http://dx.doi.org/}%
\providecommand \selectlanguage [0]{\@gobble}%
\providecommand \bibinfo  [0]{\@secondoftwo}%
\providecommand \bibfield  [0]{\@secondoftwo}%
\providecommand \translation [1]{[#1]}%
\providecommand \BibitemOpen [0]{}%
\providecommand \bibitemStop [0]{}%
\providecommand \bibitemNoStop [0]{.\EOS\space}%
\providecommand \EOS [0]{\spacefactor3000\relax}%
\providecommand \BibitemShut  [1]{\csname bibitem#1\endcsname}%
\let\auto@bib@innerbib\@empty
\bibitem [{\citenamefont {Pham}\ \emph {et~al.}(2011)\citenamefont {Pham},
  \citenamefont {Le~Sage}, \citenamefont {Stanwix}, \citenamefont {Yeung},
  \citenamefont {Glenn}, \citenamefont {Trifonov}, \citenamefont {Cappellaro},
  \citenamefont {Hemmer}, \citenamefont {Lukin}, \citenamefont {Park},
  \citenamefont {Yacoby},\ and\ \citenamefont {Walsworth}}]{Pham:2011dc}%
  \BibitemOpen
  \bibfield  {author} {\bibinfo {author} {\bibfnamefont {L.~M.}\ \bibnamefont
  {Pham}}, \bibinfo {author} {\bibfnamefont {D.}~\bibnamefont {Le~Sage}},
  \bibinfo {author} {\bibfnamefont {P.~L.}\ \bibnamefont {Stanwix}}, \bibinfo
  {author} {\bibfnamefont {T.~K.}\ \bibnamefont {Yeung}}, \bibinfo {author}
  {\bibfnamefont {D.}~\bibnamefont {Glenn}}, \bibinfo {author} {\bibfnamefont
  {A.}~\bibnamefont {Trifonov}}, \bibinfo {author} {\bibfnamefont
  {P.}~\bibnamefont {Cappellaro}}, \bibinfo {author} {\bibfnamefont {P.~R.}\
  \bibnamefont {Hemmer}}, \bibinfo {author} {\bibfnamefont {M.~D.}\
  \bibnamefont {Lukin}}, \bibinfo {author} {\bibfnamefont {H.}~\bibnamefont
  {Park}}, \bibinfo {author} {\bibfnamefont {A.}~\bibnamefont {Yacoby}}, \ and\
  \bibinfo {author} {\bibfnamefont {R.~L.}\ \bibnamefont {Walsworth}},\
  }\href@noop {} {\bibfield  {journal} {\bibinfo  {journal} {New Journal of
  Physics}\ }\textbf {\bibinfo {volume} {13}},\ \bibinfo {pages} {045021}
  (\bibinfo {year} {2011})}\BibitemShut {NoStop}%
\bibitem [{\citenamefont {Barry}\ \emph {et~al.}(2016)\citenamefont {Barry},
  \citenamefont {Turner}, \citenamefont {Schloss}, \citenamefont {Glenn},
  \citenamefont {Song}, \citenamefont {Lukin}, \citenamefont {Park},\ and\
  \citenamefont {Walsworth}}]{Barry:2016gq}%
  \BibitemOpen
  \bibfield  {author} {\bibinfo {author} {\bibfnamefont {J.~F.}\ \bibnamefont
  {Barry}}, \bibinfo {author} {\bibfnamefont {M.~J.}\ \bibnamefont {Turner}},
  \bibinfo {author} {\bibfnamefont {J.~M.}\ \bibnamefont {Schloss}}, \bibinfo
  {author} {\bibfnamefont {D.~R.}\ \bibnamefont {Glenn}}, \bibinfo {author}
  {\bibfnamefont {Y.}~\bibnamefont {Song}}, \bibinfo {author} {\bibfnamefont
  {M.~D.}\ \bibnamefont {Lukin}}, \bibinfo {author} {\bibfnamefont
  {H.}~\bibnamefont {Park}}, \ and\ \bibinfo {author} {\bibfnamefont {R.~L.}\
  \bibnamefont {Walsworth}},\ }\href@noop {} {\bibfield  {journal} {\bibinfo
  {journal} {Proceedings of the National Academy of Sciences of the United
  States of America}\ }\textbf {\bibinfo {volume} {113}},\ \bibinfo {pages}
  {14133} (\bibinfo {year} {2016})}\BibitemShut {NoStop}%
\bibitem [{\citenamefont {Glenn}\ \emph {et~al.}(2017)\citenamefont {Glenn},
  \citenamefont {Fu}, \citenamefont {Kehayias}, \citenamefont {Le~Sage},
  \citenamefont {Lima}, \citenamefont {Weiss},\ and\ \citenamefont
  {Walsworth}}]{Glenn:2017kw}%
  \BibitemOpen
  \bibfield  {author} {\bibinfo {author} {\bibfnamefont {D.~R.}\ \bibnamefont
  {Glenn}}, \bibinfo {author} {\bibfnamefont {R.~R.}\ \bibnamefont {Fu}},
  \bibinfo {author} {\bibfnamefont {P.}~\bibnamefont {Kehayias}}, \bibinfo
  {author} {\bibfnamefont {D.}~\bibnamefont {Le~Sage}}, \bibinfo {author}
  {\bibfnamefont {E.~A.}\ \bibnamefont {Lima}}, \bibinfo {author}
  {\bibfnamefont {B.~P.}\ \bibnamefont {Weiss}}, \ and\ \bibinfo {author}
  {\bibfnamefont {R.~L.}\ \bibnamefont {Walsworth}},\ }\href@noop {} {\bibfield
   {journal} {\bibinfo  {journal} {Geochemistry, Geophysics, Geosystems}\
  }\textbf {\bibinfo {volume} {18}},\ \bibinfo {pages} {3254} (\bibinfo {year}
  {2017})}\BibitemShut {NoStop}%
\bibitem [{\citenamefont {Betzig}(2015)}]{RevModPhys.87.1153}%
  \BibitemOpen
  \bibfield  {author} {\bibinfo {author} {\bibfnamefont {E.}~\bibnamefont
  {Betzig}},\ }\href {\doibase 10.1103/RevModPhys.87.1153} {\bibfield
  {journal} {\bibinfo  {journal} {Rev. Mod. Phys.}\ }\textbf {\bibinfo {volume}
  {87}},\ \bibinfo {pages} {1153} (\bibinfo {year} {2015})}\BibitemShut
  {NoStop}%
\bibitem [{\citenamefont {Bar-Gill}\ \emph {et~al.}(2013)\citenamefont
  {Bar-Gill}, \citenamefont {Pham}, \citenamefont {Jarmola}, \citenamefont
  {Budker},\ and\ \citenamefont {Walsworth}}]{BarGill:2013dq}%
  \BibitemOpen
  \bibfield  {author} {\bibinfo {author} {\bibfnamefont {N.}~\bibnamefont
  {Bar-Gill}}, \bibinfo {author} {\bibfnamefont {L.~M.}\ \bibnamefont {Pham}},
  \bibinfo {author} {\bibfnamefont {A.}~\bibnamefont {Jarmola}}, \bibinfo
  {author} {\bibfnamefont {D.}~\bibnamefont {Budker}}, \ and\ \bibinfo {author}
  {\bibfnamefont {R.~L.}\ \bibnamefont {Walsworth}},\ }\href@noop {} {\bibfield
   {journal} {\bibinfo  {journal} {Nature Communications}\ }\textbf {\bibinfo
  {volume} {4}},\ \bibinfo {pages} {ncomms2771} (\bibinfo {year}
  {2013})}\BibitemShut {NoStop}%
\end{thebibliography}%
\end{document}


\author{Soonwon~Choi}
\affiliation{Department of Physics, Harvard University, Cambridge, Massachusetts 02138, USA}
\author{Norman~Y.~Yao}
\affiliation{Department of Physics, University of California Berkeley, Berkeley, California 94720, USA}
\affiliation{Materials Science Division, Lawrence Berkeley National Laboratory, Berkeley, California 94720, USA}
\author{Mikhail~D.~Lukin}
\affiliation{Department of Physics, Harvard University, Cambridge, Massachusetts 02138, USA}

\date{\today}
\title{Supplementary Material: Quantum Metrology based on Strongly Correlated Matter}


\maketitle
\subsection{The effect of disorder on spin correlation lengths}
Our protocols introduced in the main text utilize quantum phase transitions to develop long-range spin correlations.
In an ideal case, the spin correlation length $\xi$ reaches the linear system size at the critical point, and our initialization step prepares Greenberger-Horne-Zeilinger (GHZ) states, allowing Heisenberg limited quantum sensing. 
In the presence of disorder, however, $\xi$ may be limited as the disorder can prevent the propagation of spin correlations by localizing quasi-particle excitations.
In turn, the limited spin correlations lead to diminished sensitivity enhancement compared to Heisenberg limit. 
In this section, we quantify such an effect for a spin chain with nearest neighbor Ising interactions.

For concreteness, we assume that the transverse field strength at each spin is weakly disordered $\Omega_i = \Omega + \delta \Omega_i$, where $\Omega$ is the average field strength and $\delta \Omega_i$ is a random variable uniformly distributed from $[-W/2,W/2]$. Likewise, we assume that the interaction strength $J_i = J + \delta J_i$ is also random, where $\delta J_i$ is uniformly distributed from $[-W, W]$.
Without loss of generality, the imperfection and disorder in the spin rotation angle $\delta \theta_i$ can be absorbed as a part of $\delta \Omega_i$ in the effective Hamiltonian description.

In order to quantify the localization length $\xi_\textrm{loc}$ of quasi-particles at the critical point, we numerically diagonalize our effective Hamiltonian.
More specifically, our effective Hamiltonian during the initialization step, 
$D =- \sum_i J_i S_i^z S_{i+1}^z - \sum_i \Omega_i S_i^x$,
can be mapped to a quadratic fermionic Hamiltonian
using Jorndan-Wigner transformation:
$c_i^\dagger = \left(\prod_{j=1}^{i-1} 2 S_i^x \right) (S_i^z - iS_i^y)$ and 
$c_i = \left(\prod_{j=1}^{i-1} 2 S_i^x\right) (S_i^z + iS_i^y)$. This leads to $
D= \sum_{ij}
 \left( 
\begin{array}{cc}
c_i^\dagger & c_i
\end{array}
\right) 
H_{ij}
\left(
\begin{array}{cc}
c_{j} &
 c_{j}^\dagger
\end{array}
\right)^T$, 
where the single particle Hamiltonian $H_{ij}$ can be written as 
\begin{align}
H_{ij} = 
\delta_{ij} \left(
\begin{array}{cc}
-\frac{\Omega_i}{2} & 0 \\
0 & \frac{\Omega_i}{2} 
\end{array}
\right)
 + 
 \frac{1}{8}
 \delta_{i+1,j} 
 \left(
\begin{array}{cc} 
J_i & J_i  \\
-J_i  & -J_i  
\end{array}
\right) 
 +  
 \frac{1}{8}
 \delta_{i,j+1}
 \left(
\begin{array}{cc} 
J_j & -J_j  \\
J_j  & -J_j  
\end{array}
\right).
\end{align}
We consider an even number of particles up to $N=3000$ in the even parity sector $P = 1$ with a periodic boundary condition (which corresponds to the anti-periodic boundary condition for free fermions, i.e. $c_{N+1}= -c_1$). We numerically diagonalize the single particle $H_{ij}$ at the critical point $\Omega = J/2$.

 As a proxy for an inverse localization length $\xi_\textrm{loc}^{-1}$, we compute the average inverse participation ratio $\xi_\textrm{loc}^{-1} \approx IPR = \left\langle \sum_i |\psi_i|^4 \right\rangle$, where $\psi_i$ is the quantum amplitude of an eigenstate at site-$i$,  and $\langle \cdot \rangle$ denotes averaging over 50 disorder realizations and over 50 energy eigenstates (closest to zero energy) per disorder realization. 
\begin{figure}
  \centering
  \includegraphics[width=0.38\textwidth]{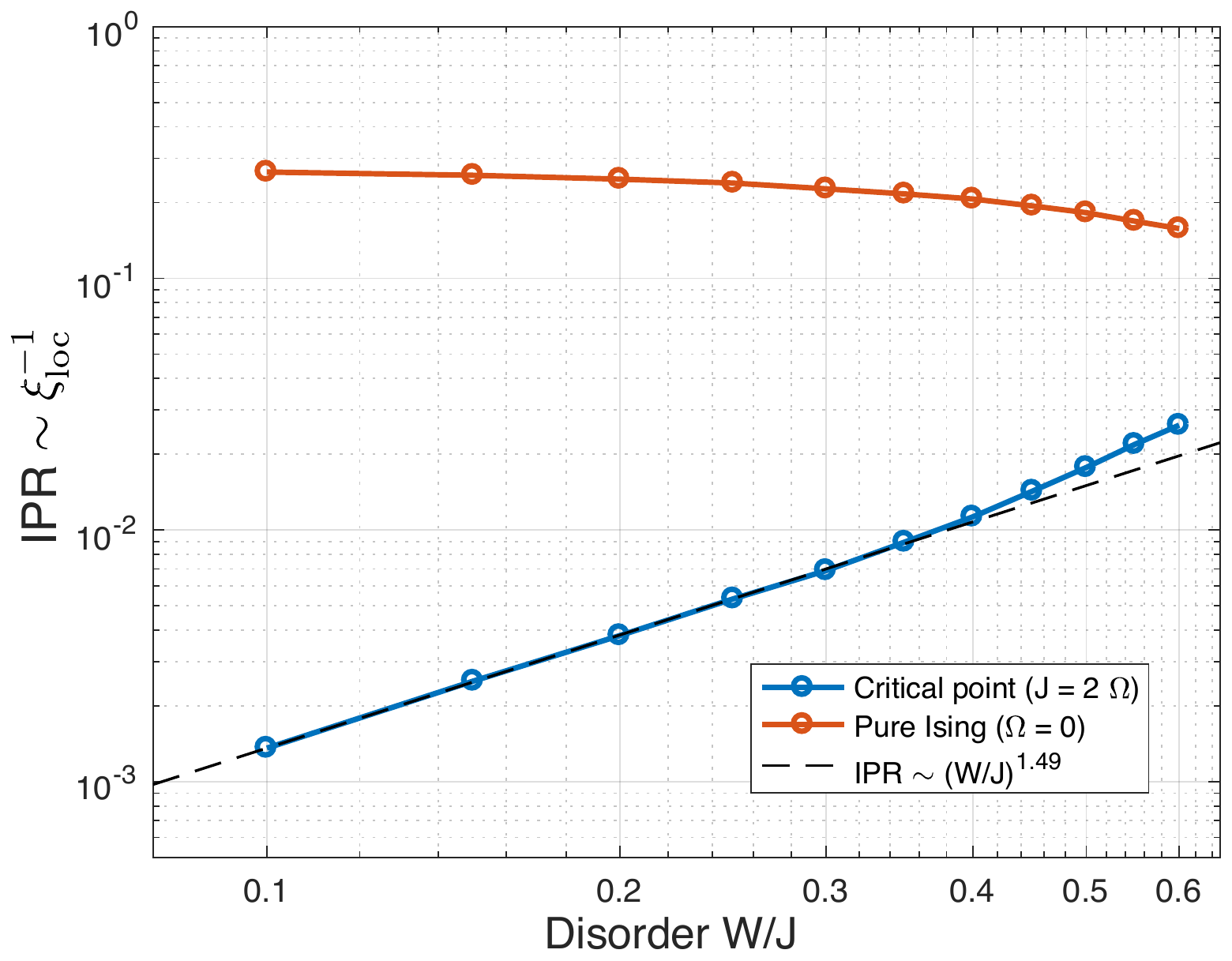}
  \caption{Inverse participation ratio as a function of disorder strength $W/J$. At the phase transition point ($\Omega = J/2$), the spatial extents of localized quasi-particle excitations are larger in orders of magnitudes than those during the measurement step ($\Omega= 0$).
  During the measurement step, the tight localization of quasi-particles is favorable for a stable accumulation of phase information.}
  \label{fig:numerics}
\end{figure}
As shown in Fig.~\ref{fig:numerics}, the localization length scales as $\xi_\textrm{loc}^{-1}\sim (W/J)^{\mu}$ with the numerically extracted exponent $\mu \approx 1.49$.

As mentioned in the main text, the effect of disorder can become favorable during the spectroscopy step.
Without disorder, domain wall excitations in one dimension are mobile, which hinders coherent accumulation of phase information over long time.
If the localization lengths during the measurement step ($\Omega = 0$) are much shorter than those at the critical point, domain wall excitations become immobile, allowing more stable accumulation of the phase information from the signal.
Repeating the numerical calculations at $\Omega = 0$, we find that the localization length is indeed multiple orders of magnitude shorter during the measurement step than at the phase transition point (Fig.~\ref{fig:numerics}).
We note that, in our numerical calculations, two types of disorder ($\delta J_i$ and $\delta \Omega_i$) are characterized by the same disorder strength $W$. However, in realistic settings, disorder in $J$ is often much stronger than that of $\Omega$, since the dominant source of disorder arises from random positioning of spins, which is even more favorable for our protocol.

\subsection{Broad-band sensing using correlated spin states}
The detection of a weak signal at an unknown frequency requires a highly sensitive spectroscopic method with a large bandwidth. Such technique is often needed in the study of fundamental physics such as the detection of gravitational waves or weakly interacting massive particles. 
In a conventional spectroscopic method, the increase in bandwidth entails the decrease in detection sensitivity. For example, as discussed in the main text, the standard quantum limit (SQL) of $N$ non-interacting particles leads to the sensitivity scaling 
\begin{align}
\delta B^{-1} \sim  \sqrt{N T T_2},
\end{align}
where $\delta B$ is the minimum detectable signal strength, $T_2$ is the duration of each measurement cycle, and $T$ is the total integration time.
The bandwidth $\delta \omega$ of this method is Fourier limited to the measurement duration $\delta \omega \sim 1/T_2$, leading to the relation between the bandwidth and the signal sensitivity
\begin{align}
\delta \omega / (\delta B)^2 \sim NT  \;\;\;\; \textrm{(conventional method in SQL)}.
\end{align}
By utilizing quantum entanglement, the bandwidth of the detection can be increased while maintaining the same signal sensitivity.
In our method, we utilize interactions to develop quantum entanglement among $\chi\gg 1$ spins (i.e. $\chi = \xi^d$ in $d$-dimensional systems),
and the sensitivity scales
\begin{align}
\label{seqn:improved_sensitivity}
(\delta B')^{-1} \sim \sqrt{N T \bar{T}_2 \chi},
\end{align}
where $\bar{T}_2$ is the relevant coherence time of the correlated spin state and can be generally shorter than the coherence time $T_2$ of individual spins. 
In particular, when each constituent spin is coupled to an independent noise bath $\bar{T}_2 \sim T_2 / \chi$, which can offset the sensitivity gain in Eq.~\eqref{seqn:improved_sensitivity}.
Nevertheless, the sensing bandwidth of our method is determined by the duration of each measurement cycle, $\delta \omega' \sim 1/\bar{T}_2$, leading to the relation
\begin{align}
\delta \omega' / (\delta B')^2 \sim \chi NT  \;\;\;\; \textrm{(quantum correlated method)}.
\end{align}
Therefore, one can improve the detection bandwidth by a factor of $\chi$ while maintaining the same sensitivity.

When our protocols are used, the number of correlated spins $\chi$ is also determined from the maximum coherence time $\bar{T}_2$ (or more precisely by $T_p \lesssim \bar{T}_2$). Since $\chi$ itself affects $\bar{T}_2$, $\chi$ has to be computed self-consistently:
\begin{align}
\chi = \xi^d &= \left( (J\bar{T}_2)^{\nu / (1 + z\nu)}\right)^d = (J T_2)^{d\nu / (1 + z \nu)}/\chi^{d\nu / (1 + z \nu)}\\
\Rightarrow & \chi^{\frac{1+z\nu + d\nu}{1 + z\nu}} = (JT_2 )^{d\nu/(1+z\nu)}\\
\Rightarrow & \chi = (JT_2)^{d\nu/(1 + z\nu + d\nu)}\label{eqn:self_consistent_chi}.
\end{align}
For example, in the case of dipolar interactions ($J_{ij} \sim J_0 /r_{ij}^3$) in a two dimensional array, the phase transition can be described by the mean-field theory with the critical exponents $\nu = 1$ and $z = 1/2$, leading to $\chi \sim \xi^2 \sim (JT_2)^{4/7}$. In a realistic setting, however, the noise bath often exhibits spatial correlations, which can modify the simple estimates provided in this section. As we discuss in the next section,  magnetic field fluctuations with microscopic origins such as dipolar spin impurities can exhibit spatial anti-correlations, which is favorable for our purposes. 

\subsection{Spatial correlation of magnetic field noise from a dipole}
When multiple spins are entangled, the coherence time $\bar{T}_2$ of the collective spin state may be different from the coherence time $T_2$ of constituent particles. More specifically, when each spin is coupled to an \emph{independent} noise bath, the coherence time of $\chi$ spins in the Greenberger-Horne-Zeilinger (GHZ) state, $\ket{G_\pm} = (\ket{\uparrow}^{\otimes \chi}\pm \ket{\downarrow}^{\otimes \chi})/\sqrt{2}$, is reduced to $\bar{T}_2 = T_2/\chi$, offsetting the potential sensitivity improvement in Eq.~(4) in the main text. 
In realistic settings, however, this simplified analysis is not valid since external noises at different spins often exhibit correlations.
Such correlations are particularly relevant in high density electronic or nuclear spin ensembles, where the decoherence arises dominantly from local sources such as interactions of sensing spins with nearby magnetic dipoles of different species.

In this section, we show that the magnetic noise generated from a fluctuating dipole (fluctuator) exhibits spatial anti-correlations at short distances. Such anti-correlations lead to a relatively longer coherence $\bar{T}_2$ compared to $T_2/\chi$, allowing sensitivity improvements for methods utilizing entangled spin states in realistic experiments. To be concrete, we begin our analysis by considering a single sensing spin interacting with a single fluctuator, which creates an effective magnetic field noise $\vec{B} (t)$ at the position of the spin. The coherence time of the sensing spin is determined from the spectral density function:  $1/T_2 \propto S^{zz}(\omega_s)$, where $\omega_s$ is the probing frequency defined in the main text, and
\begin{align}
S^{\mu \nu} (\omega) \equiv \int e^{i\omega t} \langle B^\mu (t) B^\nu (0) \rangle dt.
\label{eqn:single_spin_spectral_density}
\end{align}
When multiple spins and fluctuators are located far from one another, the noise fields at distant positions originate from different fluctuators; in such case, one may assume that spins are coupled to their own noise bath, leading to the relation $\bar{T}_2 = T_2/\chi$ discussed above.
However, when the spacing between sensing spins becomes comparable to, or even shorter than typical distances to fluctuators, the noise fields at nearby spins originate from the same fluctuator and can be correlated.
Such spatial correlations play an important role in determining the coherence time of GHZ states since the collective spins state interacts with its environment via the effective magnetic field noise
\begin{align}
\vec{B}_\textrm{eff} (t) = \sum_{i} \vec{B}(\vec{r}_i,t),
\end{align}
where $\vec{B}(\vec{r}_i,t)$ is the magnetic field experienced by a single spin positioned at $\vec{r}_i$.
The corresponding spectral density function 
\begin{align}
S_\textrm{eff}^{\mu \nu} (\omega) = \int e^{i\omega t} \sum_{ij} \langle B^{\mu} (\vec{r}_i,t) B^{\nu} (\vec{r}_j,0) \rangle dt,
\end{align}
which sensitively depends on the spatial correlations, i.e. $ \langle B^{\mu} (\vec{r}_i,t) B^{\nu} (\vec{r}_j,0) \rangle$ for $i\neq j$.
\begin{figure}
  \centering
  \includegraphics[width=0.45\textwidth]{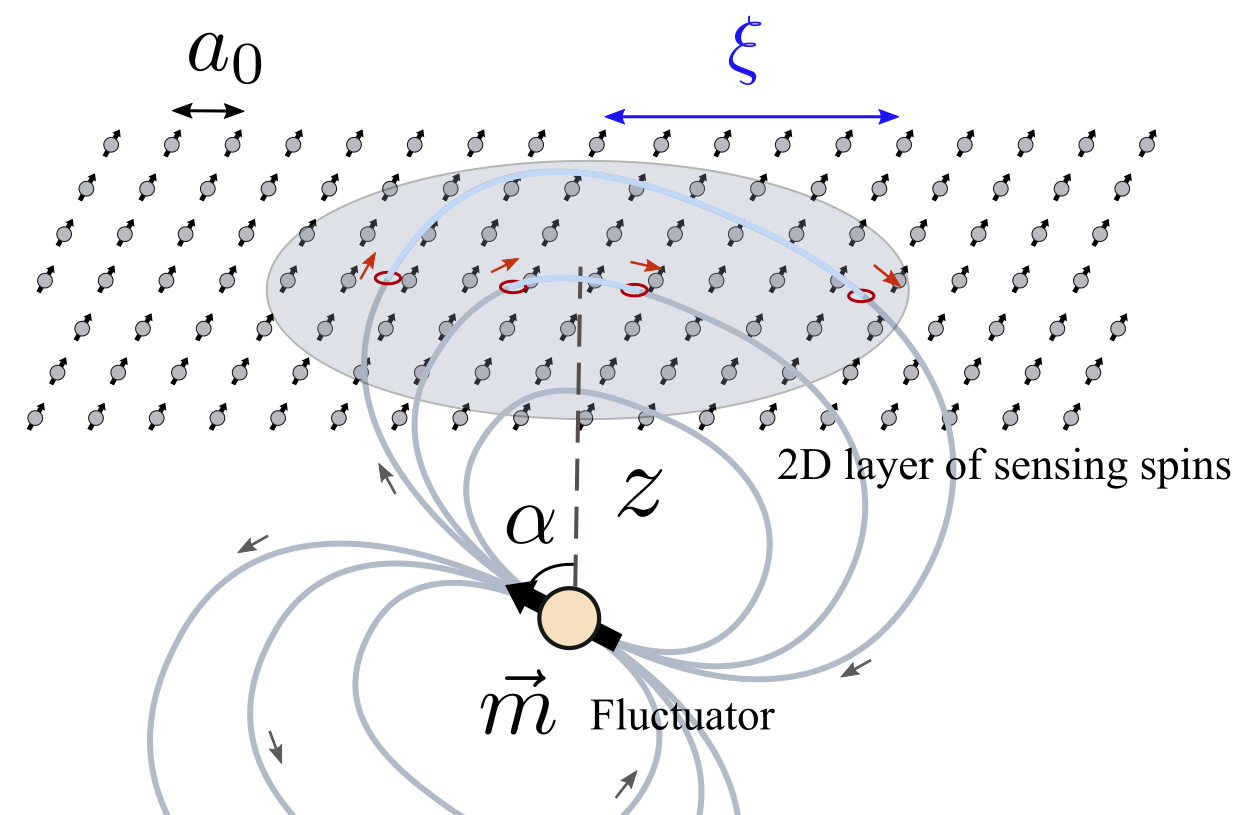}
  \caption{Schematic diagram depicting a single dipole located at distance $z$ from the sensing layer.}
  \label{sfig:spatial_corr}
\end{figure}
In order to quantify the spatial correlation, 
we consider a single magnetic dipole $\vec{m}$ located at a distance $z$ from a two-dimensional array of sensing spins with average spacing $a_0$ (see Fig.~\ref{sfig:spatial_corr}).
The fluctuator creates a classical magnetic field $\vec{B}(\vec{r}_i)$, and we compute the effective field experienced by the collective spin with correlation length $\xi$ (see Fig.~\ref{sfig:spatial_corr}):
\begin{align}
\vec{B}_\textrm{eff} &= \sum_i \vec{B}(\vec{r}_i) 
\approx \frac{1}{a_0^2} \iint \vec{B}(\vec{r}) d\vec{r}\\
&= \frac{\mu_0}{4\pi a_0^2} \iint \left( \frac{3\vec{r} (\vec{m} \cdot \vec{r})}{r^5} - \frac{\vec{m}}{r^3} \right) d\vec{r}\\
&=  \frac{\mu_0}{4 a_0^2} 
 \frac{\xi^2}{\left( z^2+ \xi^2 \right)^{3/2}}
 \left(2 m_z\hat{z}- m_q \hat{q} \right),
\end{align}
where the integration is performed over the area that covers correlated spins and $m_z = m\cos\alpha$ and $m_q = m\sin\alpha$ are the projections of the dipole moment in the vertical $\hat{z}$ and planar $\hat{q}$ directions, respectively.
%
The corresponding spectral density scales as
\begin{align}
S_\textrm{eff}^\textrm{corr} \sim \left( \frac{\mu_0 m}{a_0^2}\right)^2 \frac{\xi^4}{(z^2+\xi^2)^3}.
\end{align}
We note that when  $z \ll \xi$, the noise density $S_\textrm{eff}^\textrm{corr}$ is strongly suppressed. Intuitively, this suppression arises from the spatial profile of the magnetic field created from a dipole moment, as depicted in Fig.~\ref{sfig:spatial_corr}.
In order to obtain the total noise density coming from multiple flucuators at different depths, we need to integrate $S_\textrm{eff}^\textrm{corr}$ over all $z\in (0, \infty)$:
\begin{align}
\int_0^\infty S_\textrm{eff}^\textrm{corr} n_z d z \sim \frac{(\mu_0 m)^2}{a_0^4}  \frac{n_z}{\xi},
\end{align}
where the $n_z$ is the linear density of fluctuators along the perpendicular direction $\hat{z}$.
We neglect the integration along $\hat{x}$ and $\hat{y}$ directions due to the symmetry.
This result should be compared to a fiducial spectral density $S_\textrm{eff}^\textrm{uncorr}$ in the absence of spatial correlations, i.e. assuming $ \langle B^{\mu} (\vec{r}_i,t) B^{\nu} (\vec{r}_j,0) \rangle = 0$ for $i\neq j$.
This condition is equivalent to the assumption that each spin is coupled to an independent noise source.
For a single classical dipole at depth $z$,
\begin{align}
S_\textrm{eff}^\textrm{uncorr} \sim \sum_i |B(\vec{r}_i)|^2 \approx \frac{1}{a_0^2} \iint |\vec{B}(\vec{r})|^2 d\vec{r}
\sim \frac{(\mu_0 m)^2}{a_0^2} \left( \frac{1}{z^4} - \frac{1}{(z^2 + \xi^2)^2}\right).
\end{align}
Integrated over the entire depth $z\in (a_0, \infty)$,
\begin{align}
\int_{a_0}^\infty S_\textrm{eff}^\textrm{uncorr} n_zdz &\sim \frac{(\mu_0 m)^2}{a_0^2} \frac{n_z}{\xi^3} \int_{x=a_0/\xi}^\infty \left( \frac{1}{x^4} - \frac{1}{(1+x^2)^2}\right) dx
\sim \frac{(\mu_0 m)^2 n_z}{a_0^5} \sim \frac{ \xi}{ a_0}\int_0^\infty S_\textrm{eff}^\textrm{corr} n_z dz,
\end{align}
where we assumed that a fluctuator cannot be located closer than the spacing $a_0$.
We find that, for a sufficiently large correlation length, $\xi \gg a_0$, the integrated noise density of the correlated case is indeed smaller by a factor $a_0/\xi$ compared to the uncorrelated case.
This result implies that the effective coherence time
\begin{align}
\bar{T}_2
\sim (T_2 /\chi) ( \xi / a_0) \sim T_2 /\sqrt{\chi}.
\end{align}
Finally, from Eq.~\eqref{seqn:improved_sensitivity}, the sensitivity scales
\begin{align}
(\delta B')^{-1} &\sim \sqrt{N T \bar{T}_2 \chi}\\
\label{eqn:enhancement_correlated_noise}
&\sim \sqrt{N T T_2} \chi^{1/4}.
\end{align}
We note that this sensitivity is strictly better than SQL despite the presence of local magnetic noise sources.

\subsection{Sensitivity enhancement for magnetic field imagers}
In this section, we estimate the amount of sensitivity enhancement for magnetic field imagers when our protocols are used. A (quasi) two-dimensional array of electronic spins such as nitrogen vacancy (NV) color centers can be used to image a spatially resolved AC magnetic field profile~\cite{Pham:2011dc,Barry:2016gq,Glenn:2017kw}. 
Since quantum states of NV centers are optically read-out, the spatial resolution of this method is given by the diffraction limit of the optical wavelength $d \sim  \lambda/2 \approx 250$\,nm, and can be further improved via sub-wavelength imaging techniques~\cite{RevModPhys.87.1153}. Therefore, in order to achieve high precision AC magnetic field sensing without compromising the spatial resolution, the length scale of the probe volume should not exceed $d$.
At such length scale, however, dipolar interactions among electronic spins, $J_\textrm{dd} \sim J_0 /d^3 \approx (2\pi)\,3.3$\,Hz,  is already significant compared to the maximum coherence time of a NV center $T_2 \approx 500$\,ms at low temperature (77\,K)~\cite{BarGill:2013dq}.
 This implies that one cannot utilize more than one NV center per probe volume for the purpose of AC field sensing without affecting diffraction limited spatial resolution.
 Even at room temperature with the coherence time $T_2 \approx 3$\,ms, the separation among NV centers have to be at least $r_\textrm{min} \approx 100$\,nm in order to avoid interaction-induced decoherence. This allows one to use at most $N_0\approx 6$ NV centers per probe volume with the corresponding enhancement in signal-to-noise ratio $\sqrt{N_0} \approx 3$ at best in the conventional SQL.

Our protocols dramatically alleviate aforementioned limitations.
The minimum distance among NV centers are not bounded by dipolar interactions, allowing much higher particle density. In this case, the minimum distance is restricted only by the length scale of electronic orbitals of NV centers, which is the order of a few nm.
Assuming a single layer of NV centers, one expects to achieve $N_1 \approx 2500$ particles per probe volume. This already allows about a factor of $50$ enhancement in sensitivity at $77$\,K and about $20$ at room temperature even without accounting for the effects of quantum correlations.

\begin{figure}
  \centering
  \includegraphics[width=0.45\textwidth]{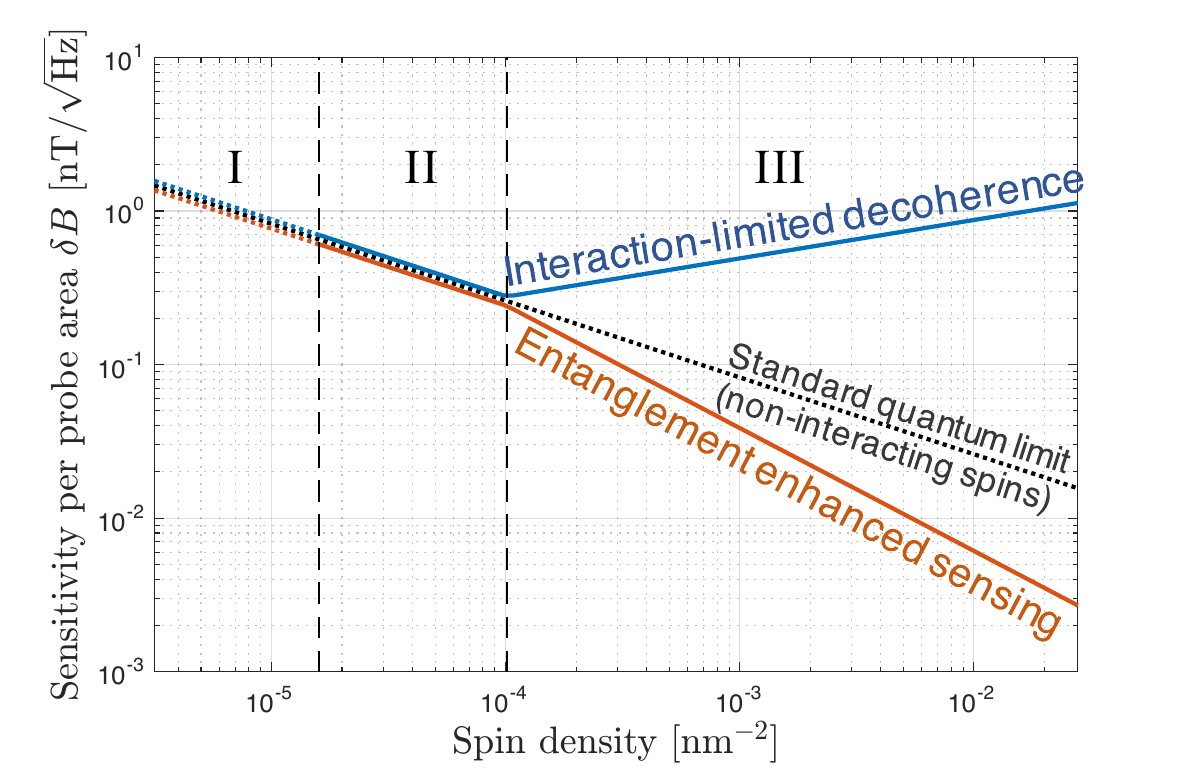}
  \caption{Sensitivity scaling of a magnetic field imager as a function of spin sensor (NV center) density. Regime I: the magnetic field imaging is limited by spin separation. Regime II and III: the spatial resolution is diffraction limited at optical wavelength $\lambda\sim$\,500\,nm. When conventional methods are used with a high density spin ensemble (regime III), the coherence time is shortened by interactions, deteriorating the sensitivity $\delta B$. In our method, such limitation is circumvented; rather, the interactions are used to further enhance the sensitivity by generating entangled many-body quantum states.
This estimate assumes two-dimensional arrays of dipolar interacting spin ensembles with a single-spin coherence time $T_2\sim 3\,$ms and the diffraction-limited probe area $\sim (\lambda/2)^2$. We assume that the dominant source of decoherence is due to local magnetic field noise created by proximal dipolar magnetic impurities. Such magnetic noises exhibit spatial correlations and allow sensitivity improvement beyond standard quantum limit when correlated spin states are used.}
  \label{fig:field_imager}
\end{figure}

When the coherence time of the sensing spin is limited by correlated magnetic noises, e.g. generated by proximal fluctuating dipoles, our protocol can provide additional sensitivity enhancement as discussed in the previous section.
In the case of two-dimensional array of spins, this sensitivity enhancement amounts to a factor of $\chi^{1/4}$ in Eq.~\eqref{eqn:enhancement_correlated_noise}. $\chi$ can be estimated self-consistently similar to Eq.~\eqref{eqn:self_consistent_chi}. Figure~\ref{fig:field_imager} summarizes the sensitivity enhancement in this scenario as a function of spin density.

\bibliography{refs}